\pdfoutput=1
\documentclass[aps,twocolumn,nobalancelastpage]{revtex4-2}
\usepackage[utf8]{inputenc}
\usepackage{microtype}
\usepackage{newtxtext}
\usepackage[upint]{newtxmath}
\usepackage{eucal}

\usepackage{graphicx}
\usepackage{booktabs}
\usepackage{enumerate}
\usepackage{siunitx}
\usepackage{braket}
\usepackage{color}
\usepackage{float}
\usepackage{soul}
\usepackage{bm}

\usepackage[colorlinks,allcolors=blue]{hyperref}
\usepackage[capitalize]{cleveref}

\usepackage[svgnames]{xcolor}
\usepackage{todonotes}
\setuptodonotes{inline}

\newcommand{\vecp}{\boldsymbol{p}}

\newcommand{\free}[1][]{\mathcal{F}_{#1}}
\newcommand{\T}{{\mathrm{T}}}
\newcommand{\dn}{\downarrow}
\newcommand{\up}{\uparrow}
\newcommand{\ph}{{\phantom{\dagger}}}
\newcommand{\ca}[1]{c^{\dagger}_{#1}}
\newcommand{\cd}[1]{c^{\vphantom{\dagger}}_{#1}}

\definecolor{Red}{rgb}{1,0,0}
\definecolor{Blue}{rgb}{0.5,0,0.9}
\definecolor{Green}{rgb}{0,1,0}

\begin{document}
\title{RKKY interaction in triplet superconductors: \\ Dzyaloshinskii--Moriya-type interaction mediated by spin-polarized Cooper pairs}
\author{Jabir Ali Ouassou}
\affiliation{Center for Quantum Spintronics, Department of Physics, Norwegian \\ University of Science and Technology, NO-7491 Trondheim, Norway}
\author{Takehito Yokoyama}
\affiliation{Department of Physics, Tokyo Institute of Technology, Meguro, Tokyo 152-8551, Japan}
\author{Jacob Linder}
\affiliation{Center for Quantum Spintronics, Department of Physics, Norwegian \\ University of Science and Technology, NO-7491 Trondheim, Norway}

\begin{abstract}
  The Ruderman--Kittel--Kasuya--Yosida (RKKY) interaction governs the coupling between localized spins and is strongly affected by the environment in which these spins reside.
  In superconductors, this interaction becomes long-ranged and provides information about the orbital symmetry of the superconducting order parameter.
  In this work, we consider the RKKY interaction between localized spins mediated by $p$-wave triplet superconductors.
  In contrast to the well-studied RKKY interaction in $d$-wave superconductors, we find that the spin of the Cooper pair in a triplet state also modulates the spin--spin coupling.
  We consider several different types of $p$-wave triplet states, and find that the form of the RKKY interaction changes significantly with the symmetries of the order parameter.
  For non-unitary superconducting states, two new terms appear in the RKKY interaction: a background spin magnetization coupling to the individual spins and, more interestingly, an effective Dzyaloshinskii--Moriya term.
  The latter term oscillates with the separation distance between the impurity spins.
  Finally, we find that the finite spin expectation value in non-unitary superconductors in concert with the conventional RKKY interaction can lead to non-collinear magnetic ground states even when the Dzyaloshinskii--Moriya term is negligible.
  The RKKY interaction in $p$-wave triplet superconductors thus offers a way to achieve new ground state spin configurations of impurity spins and simultaneously provides information about the underlying superconducting state.
\end{abstract}

\maketitle

\section{Introduction}
\begin{figure}[t]
    \includegraphics[width=\columnwidth]{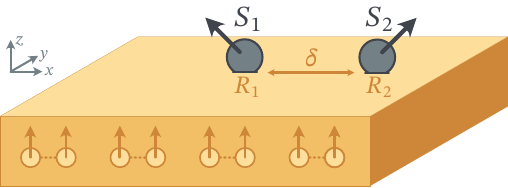}
    \caption{
        We consider a 2D superconductor in the $xy$ plane, which hosts a condensate of $p$-wave spin-triplet Cooper pairs.
        Two classical spins $\bm{S}_1$ and $\bm{S}_2$ are placed on top of this superconductor;
        $\bm{S}_1$ at the center, $\bm{S}_2$ displaced along the $x$ axis by $\delta = |\bm{R}_2 - \bm{R}_1|$.
        Herein, we investigate how the RKKY interaction between $\bm{S}_1$ and $\bm{S}_2$ is affected by unconventional superconductivity.
        Numerically, we mainly consider a square lattice of dimensions $280a\times40a$ {with open boundary conditions}, such that $\bm{R}_1 = (140a, 20a)$ and $\bm{R}_2 = (140a+\delta, 20a)$.
        Analytically, we consider {an infinite} translation-invariant {superconductor}, where only the displacement $\bm{R}_2 - \bm{R}_1$ is relevant.
        {Both the numerics and analytics show} that the RKKY interaction is sensitive to the spin and momentum symmetries of the order parameter, and can thus potentially be used to classify unconventional superconductors.
    }
    \label{fig:model}
\end{figure}

Two localized spins can interact via itinerant spin carriers in a material.
In metals, this interaction is usually mediated by electrons.
This effect is known as a Ruderman--Kittel--Kasuya--Yosida (RKKY) interaction \cite{ruderman1954a,kasuya1956a,yosida1957a} or an indirect exchange interaction.
The mechanism can be understood as follows.
When itinerant electrons approach a localized spin, the wave functions of spin-up and spin-down electrons are scattered in different ways.
This creates ``ripples'' in the net spin density, whereby the spin expectation value of the itinerant electrons oscillates and decays as a function of the distance from the localized spin.
When a second localized spin then couples to the itinerant electrons, its lowest-energy spin orientation depends on whether it has been placed in a peak or trough of the spin density generated by the first localized spin.
The net interaction between the spins can thus be either ferromagnetic or antiferromagnetic, depending on their separation distance.

The RKKY interaction is interesting for several reasons.
One is the possibility to tune the interaction, and thus the preferred alignment, of two localized spins via the system in which they are embedded.
Conversely, the behavior of the spin--spin interaction can provide important information about other interactions in the system.
Thus, the RKKY interaction has been subject of thorough investigation in a number of different classes of systems: low-dimensional electron gases \cite{yafet_prb_87, imamura2004a}, normal metals \cite{yosida1957a, ruderman1954a, kasuya1956a}, superconductors \cite{alekseevskii_zetf_77, kochelaev_zetf_79, khusainov_zetf_96, aristov1997a, bernardo_natmat_19}, and topological insulators \cite{shiranzaei_prb_17, zhu_prl_11, liu_prl_09}, to mention some examples.

When two localized spins are placed on the surface of a superconductor (see \cref{fig:model}), their RKKY interaction can provide important information about the superconducting order parameter.
A conventional BCS superconductor~\cite{bardeen1957b} has $s$-wave singlet symmetry, i.e.\ electrons experience an isotropic order parameter $\Delta(\bm{p}) = \Delta_0$ that is independent of the direction of momentum $\bm{p}$ of the electrons.
In contrast, a high-$T_c$ superconductor with $d$-wave singlet symmetry features an anisotropic gap in momentum space, e.g. $\Delta(\bm{p}) \approx (\Delta_0/p_F^2) (p_x^2 - p_y^2)$~\cite{mackenzie2003a}, where $p_F$ is the Fermi momentum.
This causes the RKKY interaction between localized spins to be highly dependent on which direction they are separated along~\cite{aristov1997a}.
Any existing spin-splitting in the superconductor makes the interaction anisotropic in spin space as well~\cite{ghanbari2021b}.
Furthermore, in non-superconducting metals, the RKKY interaction exhibits a decaying oscillation between ferromagnetic and antiferromagnetic as the separation distance~$\delta$ between the spins increases.
In singlet superconductors, an additional antiferromagnetic coupling appears~\cite{aristov1997a}.
If this coupling is sufficiently strong compared to the oscillating contribution discussed above, then the net interaction converges to purely antiferromagnetic beyond some threshold value for~$\delta$.
This superconductivity-induced coupling can also be long-ranged compared to the conventional RKKY coupling: it decays exponentially over a distance comparable to the superconducting coherence length~$\xi$, rather than exhibiting power-law decay over the Fermi wavelength~$\lambda_{\text{F}}$ as in normal metals.
Additional physics comes into play \cite{ding_pnas_21, villas_prb_21} when the RKKY interactions are modified by Yu--Shiba--Rusinov (YSR) bound states \cite{yu_aps_65, shiba_ptp_68, rusinov_zetf_69} or by the interfacial bound states that form in $d$-wave superconductors~\cite{ghanbari2021a}.

Whether the superconducting contribution to the RKKY interaction has a long range or not depends on the magnitude of~$\xi$.
Clean BCS superconductors, such as Al, can have very long coherence lengths of order 100 nm.
In contrast, high-$T_c$ cuprate superconductors typically have very short coherence lengths of order 5 nm.
In addition, the nodal structure of e.g.\ $d$-wave superconductors suppresses the superconducting contribution along some axes, since quasiparticles propagating along the nodal directions behave as in a normal metal.

Although the RKKY interaction can be directionally dependent in the presence of unconventional superconducting order, it is still spin-degenerate in singlet superconductors.
This means that the interaction between the localized spins takes the form of an effective Heisenberg interaction $\mathcal{H}_{\text{eff}} \sim \bm S_1 \cdot \bm S_2$ between the localized spins $\bm{S}_1$ and~$\bm{S}_2$.
When spin degeneracy is lifted, e.g. in spin-polarized or spin--orbit-coupled systems, new interactions emerge.
For instance, one might find Ising interactions $\mathcal{H}_{\text{eff}} \sim (\bm{S}_1 \cdot \bm{n})(\bm{S}_2 \cdot \bm{n})$ or Dzyaloshinskii--Moriya interactions $\mathcal{H}_{\text{eff}} \sim \bm{D}\cdot(\bm{S}_1\times\bm{S}_2)$, where the vectors $\bm{n}$ and~$\bm{D}$ are related to the symmetries of the underlying material.
For this reason, one would expect that spins placed on the surface of a triplet superconductor should exhibit an RKKY interaction that is anisotropic \textit{both} in direction and in spin space.
Triplet superconductors are rare, but exist: notable examples include the ferromagnetic superconductors UGe$_2$~\cite{saxena_nature_00} and URhGe~\cite{aoki_nature_01}.
Superconductors that feature spin-polarized Cooper pairs are highly sought after both for use in superconducting spintronics and for their relation to Majorana bound states~\cite{alicea_rpp_12}.
To the best of our knowledge, the RKKY interaction in triplet superconductors has not been studied so far.

In this work, we use numerical simulations to investigate the RKKY interaction between localized spins $\bm{S}_1$ and $\bm{S}_2$ on the surface of a $p$-wave triplet superconductor (see \cref{fig:model}).
In addition, we perform analytical calculations based on the Keldysh Green function formalism to verify the generality of our predictions.
Due to the rich variety of different spin and orbital structures that these unconventional superconductors can have, we consider a representative selection of different $p$-wave triplet order parameters (see \cref{tab:orders}).
We find that in unitary superconductors, the RKKY interaction contains Heisenberg and Ising contributions, whereas in non-unitary superconductors a Dzyaloshinskii--Moriya term also appears.
Moreover, the interaction is strongly sensitive to the momentum anisotropy of the $p$-wave order parameter.
Thus, we find that the RKKY interaction offers information about both the spin and momentum structure of the $p$-wave triplet state.

\begin{table}[b]
    \centering
    \caption{Superconducting order parameters considered herein.}
    \label{tab:orders}
    \begin{tabular}{lll}
        \toprule
        class                        & $d$-vector~\cite{mackenzie2003a}                             \\
        \hline
        $s$-wave singlet             & not applicable                                               \\
        $p_x$-wave triplet           & $\bm{d}(\bm{p}) = \bm{e}_z p_x$                              \\
        $p_y$-wave triplet           & $\bm{d}(\bm{p}) = \bm{e}_z p_y$                              \\
        chiral $p$-wave triplet      & $\bm{d}(\bm{p}) = \bm{e}_z (p_x + ip_y)$                     \\
        non-unitary $p$-wave triplet & $\bm{d}(\bm{p}) = (1/2) (\bm{e}_x + i\bm{e}_y) (p_x + ip_y)$ \\
        \bottomrule
    \end{tabular}
\end{table}

\section{Numerical calculation}\label{sec:numerics}
\subsection{Tight-binding model}\label{sec:tight-binding}
Consider a square lattice of dimensions $280a\times40a$, where $a$ is the lattice constant.
We take the long axis to be the $x$ axis and the short axis to be the $y$ axis.
In the absence of superconductivity, this system is described by the Hamiltonian
\begin{equation}
    \mathcal{H}_N =
    - \mu \sum_{i\sigma} \ca{i\sigma} \cd{i\sigma}
    - t \sum\limits_{\langle ij \rangle \sigma} \ca{i\sigma} \cd{j\sigma},
\end{equation}
where $t$ is the hopping and $\mu$ the chemical potential.
Henceforth, we set $\mu=-3t$.
For an explanation of how the parameters of the tight-binding model were selected, we refer to \cref{sec:params}.
The operators $\ca{i\sigma}$ and $\cd{i\sigma}$ are the usual creation and annihilation operators for spin-$\sigma$ electrons at lattice site~$i$.
{We used open boundary conditions (i.e. vacuum interfaces), meaning that the electrons cannot cross the system boundaries.}

We then place two classical spins $\bm{S}_1$ and $\bm{S}_2$ at positions $\bm{R}_1$ and $\bm{R}_2$ on the surface of the superconductor, which in the tight-binding model becomes two sites $i_1$ and $i_2$.
Each spin interacts with the itinerant electrons via an exchange interaction,
\begin{equation}
    \mathcal{H}_F = - \frac{1}{2} \mathcal{J} \sum_{i\sigma\sigma'} \sum_{p=1,2} \delta_{i,i_p} \ca{i\sigma} (\bm{S}_p\cdot \bm{\sigma})^{\vphantom{\dagger}}_{\!\sigma\sigma'} \cd{i\sigma'}.
\end{equation}
where $\bm{\sigma} = (\sigma_1, \sigma_2, \sigma_3)$ is the Pauli vector with components
\begin{align}
    \sigma_1 &=
    \begin{pmatrix}
          & \phantom{.}1\phantom{.} \\
        \phantom{.}1\phantom{.} &   \\
    \end{pmatrix}, &
    \sigma_2 &=
    \begin{pmatrix}
          & -i\phantom{.} \\
        \phantom{.}i\phantom{.} &   \\
    \end{pmatrix}, &
    \sigma_3 &=
    \begin{pmatrix}
        \phantom{.}1\phantom{.} & \\
        & -1\phantom{.}
    \end{pmatrix}.
\end{align}
Moreover, $\mathcal{J}$ is the magnitude of the spin--electron exchange coupling.
We set $\mathcal{J}=3t$ in what follows.
We normalize the spins such that $|\bm{S}_p| = 1$, which is possible since only the combined magnitude $|\mathcal{J}\bm{S}_p|$ enters the numerical model.
We place $\bm{S}_1$ at $\bm{R}_1 = (140a,20a)$ and $\bm{S}_2$ at $\bm{R}_2 = (140a+\delta, 20a)$, which for small separations $\delta$ minimizes the influence of finite-size oscillations.
One can then study the RKKY interactions in a normal metal by diagonalizing the Hamiltonian $\mathcal{H}_N + \mathcal{H}_F$ for various spin orientations~$\bm{S}_p$ and separation distances~$\delta$.

To describe superconductors, we in addition have to include the following contribution to the Hamiltonian
\begin{equation}
    \begin{aligned}
        \mathcal{H}_S = \mathcal{E}_0 - \sum_{ij\sigma} \big\{\ca{i\sigma} \big[ ({\Delta}_s \delta_{ij} + \frac12\bm{\Delta}_p \cdot \bm{\nu}_{ij})i\sigma^\ph_2 \big]^{\ph}_{\sigma\sigma'} \ca{j\sigma'} + \text{h.c.} \big\},
    \end{aligned}
\end{equation}
where $\mathcal{E}_0$ is a constant that depends on $\Delta_s$ and $\bm{\Delta}_p$.
We have here introduced the nearest-neighbor vector
\begin{equation}
    \bm{\nu}_{ij} \equiv
    \begin{cases}
        (\bm{r}_j - \bm{r}_i)/a & \text{if $|\bm{r}_j - \bm{r}_i| = a$,} \\
        \, 0                    & \text{otherwise},
    \end{cases}
\end{equation}
where $\bm{r}_i$ is the location of site~$i$.
Thus, $\Delta_s$ describes on-site $s$-wave singlet pairing, whereas $\bm{\Delta}_p$ is a $3\times2\times2$ tensor that describes off-site $p$-wave triplet pairing.
The latter is directly related to the standard $d$-vector parametrization for $p$-wave triplet order parameters~\cite{mackenzie2003a}, as shown below.
Herein, we considered both $s$-wave and $p$-wave superconducting states, and the specific $d$-vectors we used are listed in \cref{tab:orders}.
In all cases, we set the magnitude of the order parameter to $\Delta_0 = 0.1t$.

\subsection{Superconducting order parameter}
There are two common ways to describe the order parameter of a $p$-wave triplet superconductor~\cite{mackenzie2003a}.
One is a momentum-dependent matrix in spin space,
\begin{equation}
    \Delta(\bm{p}) =
    \begin{pmatrix}
        \Delta_{\up\up}(\bm{p}) & \Delta_{\up\dn}(\bm{p}) \\
        \Delta_{\dn\up}(\bm{p}) & \Delta_{\dn\dn}(\bm{p})
    \end{pmatrix},
\end{equation}
where the components enter the Fourier-transformed Hamiltonian as terms of the form $\Delta_{\sigma\sigma'}(\bm{p}) \ca{+\bm{p},\sigma} \ca{-\bm{p},\sigma'}$.
The other approach is to define a vector $\bm{d}(\bm{p})$ such that
\begin{equation}
    \Delta(\bm p) = (\Delta_0/p_F) [\bm d(\bm p) \cdot \bm{\sigma}] i\sigma_2,
\end{equation}
where $\Delta_0$ measures the overall magnitude of the order parameter.
The benefit of this approach is that the $d$-vector transforms as an ordinary vector under spin rotations.
The specific choices for $\bm{d}(\bm{p})$ considered here are listed in \cref{tab:orders}.

Numerically, we have used a third parametrization which is closely related to the above $d$-vector.
To motivate this choice, we note that any linear-in-momentum $d$-vector can be expressed in terms of a $3\times3$ tensor $\bm{D}$ such that
\begin{equation}
    \bm{d}(\bm p) = \bm{D} \bm{p}.
\end{equation}
When the $d$-vector is written as in \cref{tab:orders}, i.e.\ with all unit vectors~$\bm{e}_i$ written to the left of momentum factors $p_j$, then the corresponding $D$-tensor can be obtained by simply replacing the momentum factors $\{p_x, p_y, p_z\}$ by $\{ \bm{e}_x^\T, \bm{e}_y^\T, \bm{e}_z^\T \}$.
For example, it is straight-forward to verify that ${\bm d (\bm p) = \bm e_z (p_x + ip_y)}$ can be written $\bm d(\bm p) = \bm D \bm p$ where $\bm D = \bm e_z (\bm e_x^\T + i \bm e_y^\T)$.

Let us now return to the equation for the gap matrix $\Delta(\bm p)$.
In terms of the $D$-tensor that we introduced above, we see that $\Delta(\bm{p}) = (\Delta_0/p_F) [(\bm{D} \bm{p})^\T \bm{\sigma}] i\sigma_2 = (\Delta_0/p_F) [\bm{p}^\T \bm{D}^\T \bm{\sigma}] i\sigma_2$.
This motivates the definition of a $3 \times 2 \times 2$ tensor
\begin{equation}
    \bm{\Delta}_p \equiv \Delta_0 \, \bm{D}^\T \bm{\sigma}.
\end{equation}
In terms of this quantity, the gap matrix for a momentum~$\bm p$ can be trivially calculated as $\Delta(\bm p) = [\bm{\Delta}_p \cdot (\bm p / p_F)]i\sigma_2$.

Now, consider two nearest-neighbor sites $i, j$ on a real-space lattice.
We have previously defined a unit vector $\bm\nu_{ij}$ which points along their direction of separation.
Electrons that hop between sites $i$ and $j$ will of course have momenta $\bm{p} \sim \bm{\nu}_{ij}$, and  the real-space order parameter for two nearest-neighbor sites becomes $\Delta_{ij} = (1/2)(\bm{\Delta}_p \cdot \bm \nu_{ij})i\sigma_2$.
This form was used in the tight-binding Hamiltonian in the previous subsection, and highlights that the $D$-tensor can be a useful tool to translate general $d$-vector expressions into real-space order parameters.

\subsection{Calculating the free energy}
For a given Hamiltonian $\mathcal{H} = \mathcal{H}_N + \mathcal{H}_F + \mathcal{H}_S$ as described above, we can investigate the RKKY interaction as follows.
First, we rewrite the Hamiltonian operator $\mathcal{H}$ in terms of a $4N\times4N$ Hamiltonian matrix $\check{H}$ as
\begin{equation}
    \mathcal{H} = \mathcal{E}_0 + \frac{1}{2} \check{{c}}^\dagger \check{{H}} \check{{c}},
\end{equation}
where every creation and annihilation operator on the lattice has been collected into a single $4N$-element vector as
\begin{align}
    \check{{c}} &\equiv (\hat{c}_1, \ldots, \hat{c}_N), &
    \hat{c}_i &\equiv (\cd{i\up}, \cd{i\dn}, \ca{i\up}, \ca{i\dn}).
\end{align}
The eigenvalues $\{ \epsilon_n \}$ of $\mathcal{H}$ and $\check{H}$ are the same, but the latter is easily determined numerically.
For our calculations, we used the SciPy~\cite{virtanen2020a} function \texttt{scipy.linalg.eigh} with the \texttt{evr} driver to obtain the eigenvalue spectrum~\footnote{Performance-wise, we found the \emph{Intel Distribution for Python} to be most efficient, which ships an optimized version of SciPy that makes use of the \emph{Intel Math Kernel Library}. We also note that there was a significant performance increase by only calculating the eigenvalues (and not the corresponding eigenvectors).}.

Once the eigenvalues are known, the free energy~$\mathcal{F}$ follows from the positive eigenvalues:
\begin{equation}
    \begin{aligned}
        \mathcal{F} & = {U} - T{S}, \quad
        {U} = \mathcal{E}_0 - \frac{1}{2} \sum_{\epsilon_n > 0} \epsilon_n,            \\
        {S}         & = \sum_{\epsilon_n > 0} \log\big[ 1 + \exp(-\epsilon_n/T) \big].
    \end{aligned}
\end{equation}
This provides a way to numerically calculate the free energy $\mathcal{F}$ from the Hamiltonian $\mathcal{H}$ for each relevant spin configuration.
We consider a temperature $T = 0.001t$ far below the critical temperature~$T_c$.
As we perform non-selfconsistent calculations, $\mathcal{E}_0$ is constant and can be neglected when comparing spin configurations.
This methodology can nevertheless be used for selfconsistent calculations of the RKKY interaction as well, as reported previously for $s$-wave singlet superconductors \cite{ghanbari2021c}.

\subsection{Characterizing the spin--spin interactions}
Let us now step away from the microscopic tight-binding Hamiltonian~$\mathcal{H}$ for a moment, and consider the general free energy~$\mathcal{F}$ of a superconductor interacting with two classical spins.
    To leading order in $\bm S_1$ and $\bm S_2$, we can write
\begin{equation}
    \free(\bm S_1, \bm S_2)
    = \free[0]
        + \bm \mu_1 \cdot \bm S_1
        + \bm \mu_2 \cdot \bm S_2
        + \bm S_1 \cdot \bm \eta \bm S_2,
\end{equation}
where the $\bm \mu_i$ are vectors while $\bm \eta$ is a tensor.
We can interpret $\bm \mu_i$ as magnetic interactions between each spin and the superconductor, while $\bm \eta$ is an effective magnetic interaction between the two spins mediated by the superconductor.
The remaining $\free[0]$ contains all contributions that are independent of the spin directions.
Note that this includes e.g.\ the paramagnetic interaction between each spin and a normal metal, which usually only depends on the spin magnitude and not the spin direction.
We consider a system with a homogeneous $d$-vector, and therefore assume that $\bm \mu_1 = \bm \mu_2 \equiv \bm \mu$.
The above then becomes
\begin{equation}
    \free(\bm S_1, \bm S_2)
    = \free[0]
    + \bm \mu \cdot (\bm S_1 + \bm S_2)
    + \bm S_1 \cdot \bm \eta \bm S_2.
\end{equation}
Once we have determined the parameters $\{\free[0], \bm\mu, \bm\eta\}$, we could in principle plot the continuous function $\free(\bm S_1, \bm S_2)$ to find the ground-state spin configuration.
Note that these parameters implicitly depend on the distance~$\delta$ between the spins.

We now want a protocol to systematically extract $\bm \mu$ and $\bm \eta$ from numerical calculations of $\free(\bm S_1, \bm S_2)$.
It is convenient to define the following short-hand notation for the free energy,
\begin{equation}
    \begin{aligned}
        \free[+i,+j] & \equiv \free(+\bm{e}_i, +\bm{e}_j), &
        \free[+i,-j] & \equiv \free(+\bm{e}_i, -\bm{e}_j), \\
        \free[-i,+j] & \equiv \free(-\bm{e}_i, +\bm{e}_j), &
        \free[-i,-j] & \equiv \free(-\bm{e}_i, -\bm{e}_j),
    \end{aligned}
\end{equation}
where $\bm{e}_i, \bm{e}_j \in \{ \bm{e}_x, \bm{e}_y, \bm{e}_z \}$ are the cardinal unit vectors.
Using this notation, we see that we can write each permutation as
\begin{equation}
    \begin{aligned}
        \free[\pm i, \pm j] & = \free[0] \pm (\mu_i + \mu_j) + \eta_{ij}, \\
        \free[\pm i, \mp j] & = \free[0] \pm (\mu_i - \mu_j) - \eta_{ij}.
    \end{aligned}
\end{equation}
From these expressions, it is straight-forward to verify that we can extract all the components of $\bm \mu$ and $\bm \eta$ from
\begin{align}
    \label{eq:mu-recipe}
    \mu_i     & = \frac14 [\free[+i, +i] - \free[-i, -i]],                                \\
    \label{eq:eta-recipe}
    \eta_{ij} & = \frac14 [\free[+i, +j] - \free[+i, -j] - \free[-i, +j] + \free[-i, -j]]
\end{align}
To summarize, we have shown that if we calculate the free energy $\free(\bm{S}_1, \bm{S}_2)$ for 36 orientations $\bm{S}_1, \bm{S}_2 \in \{ \pm\bm{e}_x, \pm\bm{e}_y, \pm\bm{e}_z \}$ we can fully characterize the leading-order spin interactions.

The elements of $\bm\eta$ can be conveniently parametrized in terms of more conventional magnetic interactions between the spins,
\begin{equation}
    \begin{aligned}
        \bm{S}_1 \cdot \bm{\eta}\bm{S}_2 =\;
         & J_{x} S_{1x} S_{2x} +
        J_{y} S_{1y} S_{2y} +
        J_{z} S_{1z} S_{2z}                        \\
         & + \bm{D} \cdot (\bm S_1 \times \bm S_2)
        + \bm S_1 \cdot \bm{\Omega} \bm S_2,
    \end{aligned}
\end{equation}
where $J_n$ are exchange interactions (Heisenberg or Ising), $\bm{D}$ is an effective Dzyaloshinskii--Moriya interaction (DMI), while $\bm\Omega$ captures any remaining contributions from $\bm\eta$.
In all our simulations---including the $s$-wave and $p$-wave results presented here, and some $d$-wave and $d+is$-wave results not shown here---we have found $\bm\Omega = 0$ within numerical accuracy.
Moreover, in the analytical calculations, we have not identified any term that can not be parametrized using only $J_n$ and~$\bm{D}$.
For these reasons, we will from here on set $\bm\Omega = 0$.

Explicitly writing out each of the $J_n$ and $\bm D$ terms, we see that these can be written in terms of the following $\bm\eta$ tensor,
\begin{equation}
    \begin{aligned}
        \bm{\eta} =
        \begin{pmatrix}
            \phantom{+}J_x & +D_z           & -D_y           \\
            -D_z           & \phantom{+}J_y & +D_x           \\
            +D_y           & -D_x           & \phantom{+}J_z \\
        \end{pmatrix}.
    \end{aligned}
\end{equation}
Thus, for a system with only the contributions above, we have
\begin{align}
    J_x & = \eta_{xx}, & D_x & = \frac{1}{2} (\eta_{yz} - \eta_{zy}), \\
    J_y & = \eta_{yy}, & D_y & = \frac{1}{2} (\eta_{zx} - \eta_{xz}), \\
    J_z & = \eta_{zz}, & D_z & = \frac{1}{2} (\eta_{xy} - \eta_{yx}).
\end{align}
Thus, we can obtain the RKKY parameters $\bm{J} = (J_x, J_y, J_z)$ and $\bm{D} = (D_x, D_y, D_z)$ directly from the tensor $\bm\eta$ extracted from \cref{eq:eta-recipe}.
Notably, this parametrization lets us describe Heisenberg ($J_x = J_y = J_z$),  Ising ($J_x = J_y = 0, J_z \neq 0$), and Dzyaloshinskii--Moriya ($\bm{D} \neq 0$) interactions between spins.

\subsection{Determining the magnetic ground state}\label{sec:groundstate}
To leading order in $\bm{S}_1$ and $\bm{S}_2$, we have shown that the free energy of the system in \cref{fig:model} can be characterized as
\begin{equation}
    \label{eq:free-energy-optim}
    \mathcal{F} = \mathcal{F}_0 + \bm{\mu}\cdot(\bm{S}_1 + \bm{S}_2) + \bm{J}\cdot(\bm{S}_1 \circ \bm{S}_2) + \bm{D} \cdot (\bm{S}_1 \times \bm{S}_2),
\end{equation}
where $\circ$ refers to elementwise multiplication of two vectors.
Moreover, we have demonstrated how to calculate the parameters of this free energy from numerical simulations.
Once these are known, we can determine the preferred orientations of the spins by minimizing $\mathcal{F}$ with respect to $\bm{S}_1$ and $\bm{S}_2$.
In real physical systems, there will likely be additional contributions to the free energy due to e.g.\ magnetocrystalline anisotropy.
Such contributions depend on the specific materials and geometries under consideration, and have been neglected in this study.
Moreover, the calculation procedure for $\{\bm\mu, \bm J, \bm D\}$ assumes that only linear and bilinear terms in $\bm{S}_1$ and $\bm{S}_2$ exist in the free energy.
Higher-order terms in the free energy could therefore affect the numerically obtained values for these parameters.

The mathematical problem we need to solve is a constrained optimization problem.
Specifically, we need to minimize $\mathcal{F}$ while ensuring that $\bm{S}_1$ and $\bm{S}_2$ remain unit vectors:
\begin{equation}
    \begin{aligned}
        &\text{minimize} & &\mathcal{F}(\bm{S}_1, \bm{S}_2) \\
        &\text{subject to} & &|\bm{S}_1| = |\bm{S}_2| = 1.
    \end{aligned}
\end{equation}
To solve this numerically, we first convert it to an unconstrained optimization problem.
This can done by incorporating the constraints into the objective function as penalty terms:
\begin{equation}
    \mathcal{G}(\bm{S}_1, \bm{S}_2) = \mathcal{F}(\bm{S}_1, \bm{S}_2) + \lambda |\bm{S}_1^2 - 1|^2 + \lambda |\bm{S}_2^2 - 1|^2.
\end{equation}
For suitably large penalties~$\lambda$, the constraints are automatically satisfied as the penalized objective function $\mathcal{G}$ is minimized.
As part of the optimization procedure described below, we will therefore numerically let $\lambda \rightarrow \infty$ to enforce these constraints.

Non-convex optimization requires an initial guess for the variables $\bm{S}_1$ and $\bm{S}_2$, and can identify different local minima in $\mathcal{F}$ depending on this guess.
To find the \emph{global} energy minimum (or minima), we therefore have to repeat the optimization procedure with different starting points and compare the final results.
In our case, we used as initial guesses
\begin{equation}
    \bm{S}_{1}, \bm{S}_{2} \in \Big\{ \cos \theta\, \bm{e}_x + \sin \theta\, \bm{e}_z \;\Big|\; \theta \in \{0^\circ, 45^\circ, \ldots, 360^\circ \} \Big\},
\end{equation}
which produces uniformly spaced starting points in the $xz$ plane.
In our study, it is sufficient to focus on the $xz$ plane since all $d$-vectors listed in \cref{tab:orders} were found to produce $J_x = J_y$, $\bm{\mu} \sim \bm{e}_z$, and $\bm{D} \sim \bm{e}_z$.
Under these conditions, the free energy is invariant under spin rotations in the $xy$ plane.

For each initial guess, we minimized $\mathcal{G}(\bm{S}_1, \bm{S}_2)$ using the SciPy~\cite{virtanen2020a} function \texttt{scipy.optimize.minimize} with the Powell backend.
The penalty $\lambda \in \{10^0, 10^1, \ldots, 10^6\}$ was increased exponentially between sucessive calls to the optimizer; this eased the convergence of initial iterations, but strongly enforced the constraints in final iterations.
After the final call to the optimizer, we saved the optimized spin configuration $(\bm{S}_1, \bm{S}_2)$ and free energy $\mathcal{F}(\bm{S}_1, \bm{S}_2)$.
After an independent optimization for each initial guess, the real ground state was taken to be the optimization result with the lowest value for $\mathcal{F}$.

We have verified that for systems with known ground states (e.g.\ systems with pure Heisenberg, Ising, or Dzyaloshinskii--Moriya interactions), the procedure above reproduces the known ground states.
The procedure was then applied to the numerically calculated $\mathcal{F}(\bm{S}_1, \bm{S}_2)$ for $p$-wave triplet superconductors with various separations $\delta$ between the spins.

\subsection{Parameter study}\label{sec:params}
When a spin $\bm{S}$ is placed on the surface of a superconductor, a Yu--Shiba--Rusinov (YSR) bound state can emerge~\cite{yu_aps_65, shiba_ptp_68, rusinov_zetf_69}.
This state manifests as a peak within the subgap local density of states (LDOS).
The amplitude of this peak oscillates and decays with increased distance from the spin.
When these bound states (i)~have a high amplitude and (ii)~are visible farther from the spin site, we can infer that the spin must more strongly affect the material on which it is placed.
Intuitively, we would also expect RKKY interactions between spins to be enhanced in this limit.
It is much more computationally efficient to look for YSR signatures (LDOS calculations) than to determine RKKY interactions (free energy calculations).
We therefore used YSR signatures as a proxy to determine optimal parameters for our RKKY calculations.
Our methodology for efficient LDOS computation on large lattices is detailed in \cref{sec:ldos}.

We set the superconducting gap $\Delta_0 = 0.1t$.
Lower values would be more realistic for typical low-temperature super\-conductors;
however, too small values make it challenging to pick out the YSR state amid subgap peaks caused by finite size effects.
One possible remedy is to consider extremely large lattices, thus abating the finite size effects.
However, the required system sizes quickly become computationally infeasible.
Our choice $\Delta_0 = 0.1t$ strikes a balance between realism and feasiblity, as is common in numerical treatments of superconductivity based on tight-binding models.

For the exchange coupling between each classical spin and the superconductor, a moderately large value $\mathcal{J} = 3t$ was chosen.
A significant ratio $\mathcal{J}/\Delta_0 = 30$ as chosen here is not uncommon; for instance, Ref.~\cite{yazdani1997a} estimated a ratio of over a thousand for Mn adatoms on an Nb superconductor.
In practice, this parameter $\mathcal{J}$ can vary significantly depending on the specific materials used in an experiment, and the YSR signatures become clearer the larger this parameter grows.
Notably, some experimental setups also permit in-situ tuning of its precise value~\cite{farinacci2018a}.

Lastly, a chemical potential $\mu \approx -3t$ was deemed optimal for obtaining pronounced YSR signatures.
As $\mu \rightarrow -4t$ the amplitude of the YSR state becomes negligible, whereas for $\mu \rightarrow 0$ the YSR state becomes so strongly localized that it is only clearly visible from the nearest-neighbor site.
The value $\mu = -3t$ strikes a good compromise, where a clear LDOS subgap peak is visible from several lattice sites away.

After obtaining $\{\Delta_0 = 0.1t, J = 3t, \mu = -3t\}$ from studying YSR states in $p$-wave superconductors, we proceeded to study RKKY interactions with the same parameters as a basis.
We believe that the general conclusions of our study are robust the specific system parameters---especially since the same qualitative contributions were also derived analytically in \cref{sec:analytics} without reference to these particular system parameters.

\section{Analytical calculation}\label{sec:analytics}
Consider a homogeneous {superconductor} described by a $4\times4$ Green function $\hat{G}^R(\bm{p}, \omega)$ in Nambu$\otimes$spin space, where $\bm{p}$ and $\omega$ are momentum and energy variables.
As in the numerical case, we then place two classical spins $\bm{S}_{1,2}$ at coordinates~$\bm{R}_{1,2}$, where each spin couples to the {superconductor} via local exchange interactions with strength~$\mathcal{J}$.
Leading-order perturbation theory then shows that the equilibrium energy of the system acquires a term
\begin{equation}
    \begin{aligned}
        E_{\text{RKKY}} = &                                      \\
        {\frac{1}{4} \pi} \mathcal{J}^2 \, \mathrm{Im}
                          & \int d\omega\, \tanh(\omega/2T)      \\
        \times\,          & \int d\bm{p}_{1} \int d\bm{p}_{2} \,
        e^{{-i}(\bm{p}_{2} - \bm{p}_{1})\cdot(\bm{R}_2 - \bm{R}_1)} \\
        \times\,          & \mathrm{Tr}\,
        \Big\{\,
        (\bm{S}_1 \cdot \hat{\bm{\sigma}})
        \, \hat{G}^R(\bm{p}_1, \omega)
        \,(\bm{S}_2\cdot\hat{\bm{\sigma}})
        \, \hat{G}^R(\bm{p}_2, \omega)
        \,\Big\},
    \end{aligned}
\end{equation}
where $\hat{\bm{\sigma}} = \text{diag}(\bm{\sigma}, \bm{\sigma}^*)$.
Although this is a well-established equation in the literature (see e.g.\ Refs.~\cite{schwabe1996a,imamura2004a,wang2017a}), we provide a complete derivation in \cref{sec:general}.
The RKKY interaction can then be understood as a reorientation of $\bm{S}_1$ and $\bm{S}_2$ in an attempt to minimize this term in the system's energy.

Consider now a $p$-wave triplet superconductor.
The Nambu-space structure of its retarded Green function can be written
\begin{equation}
    \hat{G}^R(\bm{p}, \omega) = \begin{pmatrix}
        G(\bm{p}, \omega)         & F(\bm{p}, \omega)         \\
        \tilde{F}(\bm{p}, \omega) & \tilde{G}(\bm{p}, \omega) \\
    \end{pmatrix},
\end{equation}
where the normal component~$G$ describes quasiparticles, the anomalous component~$F$ describes Cooper pairs, and $\tilde{X}(\bm{p}, \omega) \equiv X^*(-\bm{p}, -\omega)$.
Substituting this into the equation for $E_{\text{RKKY}}$, one can show that (see \cref{sec:specific})
\begin{equation}
    \begin{aligned}
        E_{\text{RKKY}} =
        {\frac{1}{2} \pi} \mathcal{J}^2 \, \mathrm{Im}
               & \int d\bm{p}_{1} \int d\bm{p}_{2} \,
        e^{{-i}(\bm{p}_{2} - \bm{p}_{1})\cdot(\bm{R}_2 - \bm{R}_1)} \\
        \times & \int d\omega\, \tanh(\omega/2T) \,
        (\mathcal{G} + \mathcal{F}),
    \end{aligned}
    \label{eq:rkky-in-paper}
\end{equation}
where the contributions from quasiparticles and Cooper pairs are given by the two amplitudes
\begin{align}
    \mathcal{G} & = \mathrm{Tr}[(\bm{S}_1\cdot\bm{\sigma}) G(\bm{p}_1, \omega) (\bm{S}_2\cdot\bm{\sigma}) G(\bm{p}_2, \omega)],           \label{eq:rkky-g-in-paper} \\
    \mathcal{F} & = \mathrm{Tr}[(\bm{S}_1\cdot\bm{\sigma}) F(\bm{p}_1, \omega) (\bm{S}_2\cdot\bm{\sigma}^*) \tilde{F}(\bm{p}_2, \omega)]. \label{eq:rkky-f-in-paper}
\end{align}
In non-superconducting metals, $\mathcal{G}$ typically gives rise to a Heisenberg interaction $\sim \bm{S}_1 \cdot \bm{S}_2$ between the two spins~\cite{ruderman1954a,kasuya1956a,yosida1957a}, while $\mathcal{F} = 0$ in the absence of superconductivity.
In spin--orbit-coupled materials, $\mathcal{G}$ can also give rise to Ising $\sim (\bm{S}_1\cdot\bm{n})(\bm{S}_2\cdot\bm{n})$ and Dzyaloshinskii--Moriya $\sim\bm{D}\cdot(\bm{S}_1\times\bm{S}_2)$ interactions~\cite{imamura2004a,wang2017a}.

  To rigorously describe the RKKY interaction below the superconducting critical temperature~$T_c$ we need to evaluate \cref{eq:rkky-in-paper,eq:rkky-g-in-paper,eq:rkky-f-in-paper}.
  In \cref{sec:green-explicit}, we consider a $p$-wave triplet superconductor described by a general $d$-vector $\bm{d}(\bm{p})$.
  We then explicitly calculate the Green function via block-matrix inversion, and find that the result can be written in the form
  \begin{align}
    G(\bm p, \omega) &= g_s(\bm p, \omega) + \bm{g}_p(\bm p, \omega) \cdot \bm\sigma, \\
    F(\bm p, \omega) &= [\bm{f}_p(\bm p, \omega) \cdot \bm\sigma] i\sigma_2,
  \end{align}
  where $g_s$ is a scalar, $\bm{g}_p \sim \bm{d} \times \bm{d}^*$ is proportional to the spin expectation value of the triplet condensate, and $\bm{f}_p \sim \bm{d}$ is proportional to the $d$-vector that describes the condensate.

  In \cref{sec:specific}, we then use the results above to calculate the quasiparticle contribution~$\mathcal{G}$ and condensate contribution~$\mathcal{F}$.
  The final result of this derivation can be written as follows:
  \begin{equation}
    \begin{aligned}
      E_{\text{RKKY}} \sim \mathcal{J}^2 \int d\omega\, &\tanh(\omega/2T) \\
      \times \;\mathrm{Im} \,  & \Big\{\,(g_{s}^2 - \bm{g}_{p}^2 - \bm{f}_{p} \cdot \tilde{\bm{f}}_{p}) (\bm{S}_1 \cdot \bm{S}_2) \\
                                   &+ 2(\bm{f}_{p} \cdot \bm{S}_1) (\tilde{\bm{f}}_{p} \cdot \bm{S}_2) \\
                                   &+ 2(\bm{g}_{p} \cdot \bm{S}_1) (\bm{g}_{p} \cdot \bm{S}_2) \\
                                   &+ (\bm{f}_p \times \tilde{\bm{f}}_p) \cdot (\bm{S}_1 \times \bm{S}_2) \Big\}.
    \end{aligned}
    \label{eq:rkky-in-paper-final-version}
  \end{equation}
  Here, $\{ g_s(\bm{R}_2 - \bm{R}_1, \omega), \bm{g}_p(\bm{R}_2 - \bm{R}_1, \omega), \bm{f}_p(\bm{R}_2 - \bm{R}_1, \omega) \}$ are real-space Green functions evaluated at the distance $\bm{r} = \bm{R}_2 - \bm{R}_1$ between the spins.
  These real-space Green functions are naturally obtained from the momentum-space Green functions defined above via a Fourier transform (see \cref{sec:specific}).

  In normal metals only $g_s \neq 0$, so we find a pure Heisenberg interaction between the spins: $E_{\text{RKKY}} \sim \bm{S}_1\cdot\bm{S}_2$.
  However, this term is sensitive to the presence of superconductivity, as this activates additional contributions like $\bm{f}_p \cdot \tilde{\bm{f}}_p \sim \bm{d} \cdot \bm{d}^* \sim |\Delta|^2$.
  This can be understood as follows.
  Superconductivity opens a directionally dependent gap $|\Delta(\bm{p})|$ in the quasiparticle excitation spectrum.
  This strongly affects the mobility of the system's quasiparticles in the gapped directions, thus modulating the interactions that these quasiparticles can mediate.
  Similar contributions appear in both $s$-wave and $d$-wave singlet superconductors~\cite{aristov1997a}.
However, the momentum symmetry differs for $p$-wave superconductors as
  $\bm{f}_p(\bm{R}_2 - \bm{R}_1, \omega)$ is a highly anisoptropic function of the spin separation direction~\footnote{This follows because $\bm{f}_p(\bm{r}, \omega)$ is the Fourier transform of $\bm{f}_p(\bm{p}, \omega) \sim \bm{d}(\bm{p})$. Since the $d$-vector is highly anisotropic, $\bm{f}_p(\bm{R}_2 - \bm{R}_1, \omega)$ must be highly anisotropic as well.}.

  In the presence of $p$-wave triplet superconductivity, we also find an Ising interaction between $\bm{S}_1$ and $\bm{S}_2$.
  In this case, only
spin components along some special axis $\bm{n}$ couple: $E_{\text{RKKY}} \sim (\bm{S}_1\cdot\bm{n})(\bm{S}_2\cdot\bm{n})$.
For a concrete example, consider a $p_x$-wave superconductor: $\bm{d}(\bm{p}) \sim p_x \bm{e}_z$.
The Ising term will then be proportional to $(\bm{S}_1\cdot\bm{e}_z)(\bm{S}_2\cdot\bm{e}_z) = S_{1z} S_{2z}$, such that only the $z$ components of the two spins couple.
Naturally, as this term is also proportional to $\bm{f}_p$ and $\tilde{\bm{f}}_p$, it also causes a strongly anisotropic coupling between the spins.

  Next, consider the third term in the integrand of \cref{eq:rkky-in-paper-final-version}.
  This is an Ising interaction along the spin expectation value $\bm{d} \times \bm{d}^*$ of the superconducting condensate.
  This term differs from the previous Ising term in two key ways.
  Firstly, the previous term should arise for \emph{any} $p$-wave triplet superconductor ($\bm{d} \neq 0$), whereas the current term would arise only for \emph{non-unitary} superconductors ($\bm{d} \times \bm{d}^* \neq 0$)~\cite{mackenzie2003a}.
  Secondly, when both contributions exist, the two Ising contributions tend to order spins along perpendicular axes: $\bm{g}_p \sim \bm{d} \times \bm{d}^* \sim \bm{f}_p \times \tilde{\bm{f}}_p$.

  Finally, the detailed analytical derivation in \cref{sec:specific} also uncovered a fourth and qualitatively distinct contribution to the RKKY interaction:
  Namely, an effective Dzyaloshinskii--Moriya interaction  $E_{\text{RKKY}} \sim \bm{D}\cdot(\bm{S}_1\times\bm{S}_2)$~\cite{dzyaloshinsky1958a,moriya1960a}.
  The resulting DMI vector is proportional to $\bm{f}_p \times \tilde{\bm{f}}_p \sim \bm{d} \times \bm{d}^*$, and thus appears to be another contribution specific to non-unitary superconductors.
  However, upon further analysis (see \cref{sec:specific}), it was found that for infinite translation-invariant superconductors, the DMI vector $\bm{D} = 0$.
  As we will see in \cref{sec:results-disc}, our numerical simulations nevertheless show that such a DMI contribution does arise in \emph{finite superconductors with broken inversion symmetry due to edges}---even in the absence of spin-orbit coupling in the model.
  Moreover, consistent with the analytical expression, the numerical results show that the DMI vector is an $\mathcal{O}(\mathcal{J}^2)$ term that only appears for $\bm{d} \times \bm{d}^* \neq 0$.

The results above show that the RKKY interaction in $p$-wave triplet superconductors depend sensitively on the spin symmetries of the triplet superconducting order parameter.
In addition, all these interactions depend on the momentum anisotropy in $\bm{d}(\bm{p})$ via the Green functions $\{ g_s, \bm{g}_p, \bm{f}_p \}$.
This shows that the RKKY interaction can be used as a probe for both the spin and momentum symmetries of the underlying $p$-wave triplet order parameter.
As we will see in the next section, these analytical predictions fit well with the numerical results obtained using the methodology presented above.

  Some of the RKKY contributions above only arise in non-unitary triplet superconductors.
  However, we note that one does not necessarily require an intrinsic non-unitary triplet superconductor to access such a quantum state.
This type of superconductivity is also known to arise in hybrid structures comprised of ferromagnets and conventional superconductors \cite{buzdin_rmp_05, bergeret_rmp_05, linder_nphys_15, eschrig_rpp_15}.
This could then serve as an alternative, and perhaps experimentally more accessible, arena for probing the RKKY interaction mediated by a non-unitary superconducting state.

\section{Results and discussion}
\label{sec:results-disc}
\begin{figure*}
    \includegraphics[width=\textwidth]{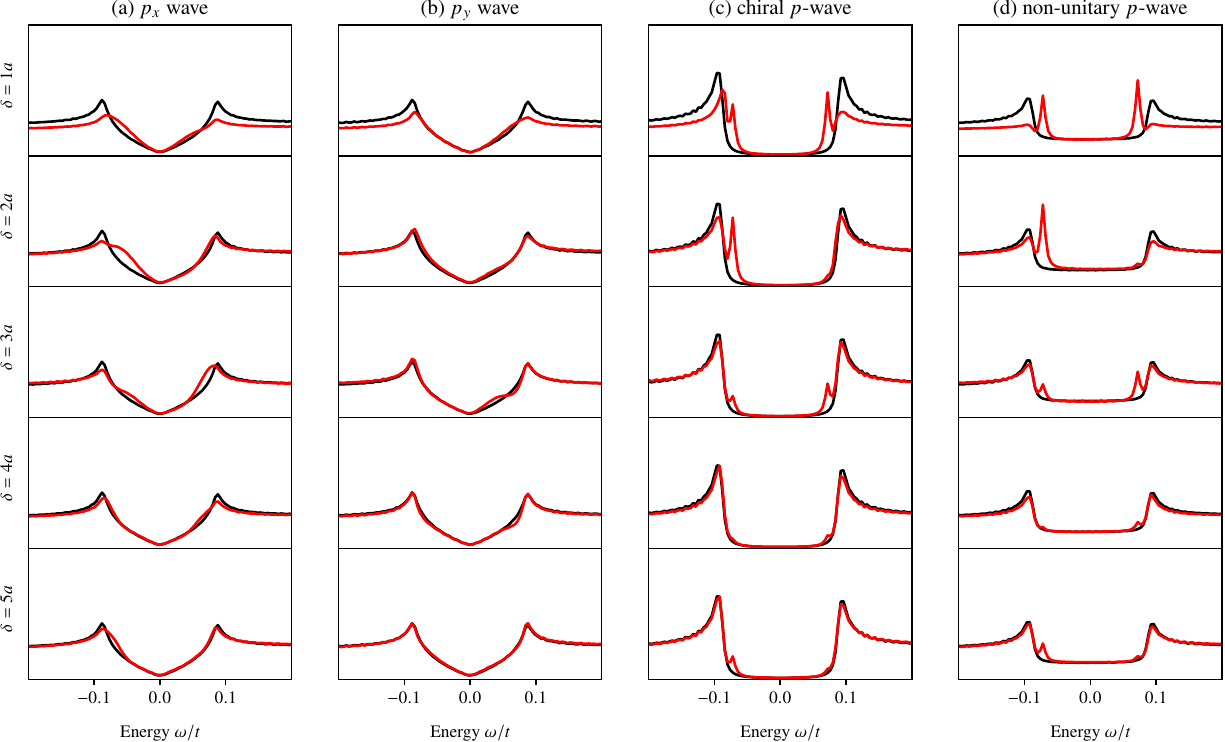}
    \caption{
            Numerical results for the LDOS in various $p$-wave superconductors, where the specific order parameters are indicated in the column headers (cf. \cref{tab:orders}).
            Each row shows plots for a different distance $\delta$ along the $x$ axis from the center of the lattice where a spin impurity has been placed.
            The red lines show the LDOS with a spin impurity in the system, whereas the black lines show the corresponding LDOS in the absence of any spin impurities for comparison.
            The YSR states are thus visible as subgap peaks that arise in the red but not black line plots.
            This effect is perhaps most pronounced for the chiral $p$-wave, but it can also be seen as an enlargement of the subgap LDOS of e.g. the $p_x$ wave state.
    }
    \label{fig:ysr}
\end{figure*}

    As discussed in \cref{sec:params}, it is numerically much more efficient to calculate the LDOS (cf. \cref{sec:ldos}) than to calculate the free energy.
    For this reason, we decided to use YSR states as a guide to parameter optimization before proceeding to calculate the RKKY interaction.
      To investigate this effect, we considered $N \times N$ square lattices where a spin impurity had been placed at the center $(N/2, N/2)$ of the system.
      We then calculated the LDOS at a site $(N/2 + \delta, N/2)$ that is located a distance $\delta$ from the spin impurity along the $x$ axis.
      In the presence of YSR bound states, we should then find subgap ``spikes'' in these LDOS plots, where the height of these spikes oscillates and decays as a function of~$\delta$.
    The numerical results for the LDOS near a spin impurity is shown in \cref{fig:ysr} for the final model parameters $\{ \Delta_0 = 0.1t, J = 3t, \mu = -3t \}$ that we obtained through this procedure.
    The plots in \cref{fig:ysr} were calculated on $800\times800$ lattices; for small lattice sizes, the results are qualitatively similar, but many smaller spikes also appear in the LDOS due to finite size effects.

    It is worth noting that the YSR signatures in \cref{fig:ysr}(a) are significantly more smeared out compared to the clean spikes that appear in \cref{fig:ysr}(c--d).
    This is most likely due to a hybridization between the YSR states and the subgap states that already exist in a $p_x$-wave superconductor in the absence of the spin impurity.
    Nevertheless, when compared to the black lines which show the LDOS in the absence of a spin impurity, the asymmetric enhancement of the subgap LDOS due to the presence of YSR states is quite clear.
    Note also the comparatively small LDOS changes that are visible in \cref{fig:ysr}(b).
    Since we consider a displacement along the $x$ axis from the location of the spin impurity, and a $p_y$-wave state primarily has superconducting properties along the $y$ axis, this is not so surprising.

\begin{figure}[b!]
    \includegraphics[width=\columnwidth]{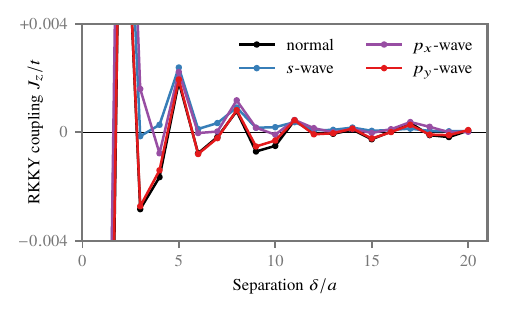}
    \caption{
        Comparison of the RKKY coupling $J_z$ for the simplest  $p$-wave order parameters with normal metals and $s$-wave superconductors.
        Note that the results for the $p_x$-wave order parameter {are} qualitatively similar to {those for} the $s$-wave order parameter, whereas {the results for} the $p_y$-wave order parameter {are} nearly indistinguishable from the normal metal results.
    }
    \label{fig:baseline}
\end{figure}

\begin{figure*}
    \includegraphics[width=\textwidth]{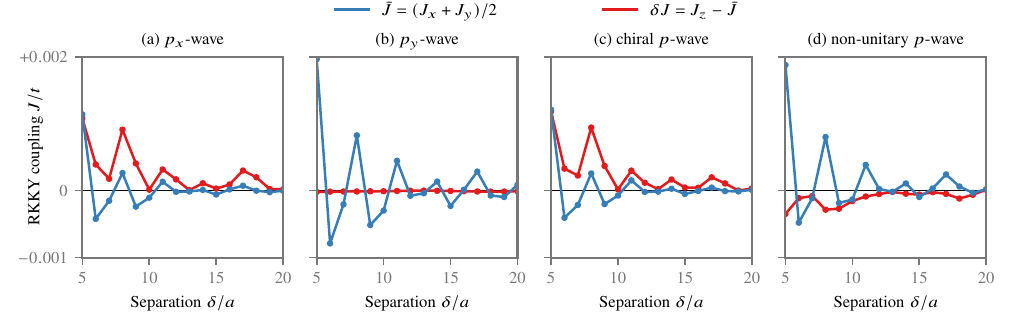}
    \includegraphics[width=\textwidth]{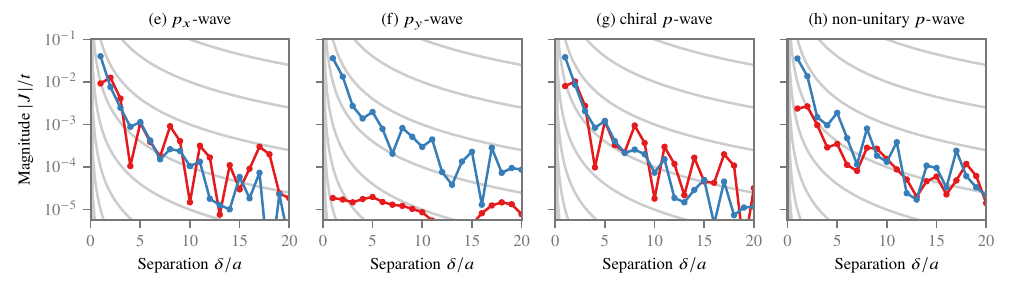}
    \caption{
        (a--d)~Comparison of the different RKKY interaction components $\{ J_x, J_y, J_z \}$ for the considered $p$-wave order parameters.
        The results are clearly sensitive to the spin part of the $\bm{d}$ vector: e.g.\ for $\bm{d} = \bm{e}_z p_x$ the interaction is clearly more antiferromagnetic for $J_z$ than for $J_x, J_y$.
        In the case of the non-unitary superconductor, the interaction is more antiferromagnetic for \emph{both} the $x$ and $y$ directions as compared to the $z$ direction, which in our parametrization leads to a negative value for $\delta J$.
            (e--h)~Plots of the magnitude $|\bar{J}|$ and $|\delta J|$ of the same data shown on a logarithmic scale.
            In the background, we have added helper lines $10^{-n} / \delta^2$ for various integers $n$ as grey lines.
            This suggests that the amplitude of the RKKY oscillations in $p$-wave superconductors decay as roughly $\sim 1/\delta^2$, consistent with previous work on 2D RKKY interactions \cite{aristov1997a}.
    }
    \label{fig:comparison}
\end{figure*}

The simplest $p$-wave triplet state is arguably the $p_x$-wave state described by $\bm{d}(\bm{p}) = p_x \bm{e}_z$.
In this state, the super\-conducting gap is highly anisotropic in momentum space:
\begin{align}
    |\Delta(\bm{p} = \pm p_F \bm{e}_x)| &= \Delta_0, &
    |\Delta(\bm{p} = \pm p_F \bm{e}_y)| &= 0.
\end{align}
Thus, when two spins $\bm{S}_1$~and~$\bm{S}_2$ are displaced along the $x$ axis on the surface of this superconductor (see \cref{fig:model}), quasiparticles traveling directly from one spin to the other would ``see'' a full gap at the Fermi level.
This situation is qualitatively similar to in an $s$-wave superconductor.
On the other hand, if the spins are displaced along the $y$ axis, quasiparticles would see no gap and thus behave similarly to in a normal metal.
In this case, rotating the whole physical system by $90^\circ$ in the $xy$ plane would yield a $p_y$ wave superconductor with spins separated along the $x$ axis.
We have opted to study the latter case in this paper.

Using the methodology presented in \cref{sec:numerics}, we have numerically calculated the RKKY coupling $J_z$ between two spins when placed on $p_x$ and $p_y$ wave superconductors.
This coupling corresponds to an effective interaction $J_z S_{1z} S_{2z}$ in the free energy of the system.
In \cref{fig:baseline}, these results are compared to the corresponding RKKY interactions in normal metals and $s$-wave superconductors.
In line with the argument above, we find that the results for $s$-wave and $p_x$-wave superconductors are quite similar, whereas the results for normal metals and $p_y$-wave superconductors are nearly indistinguishable.

In normal metals and singlet superconductors, the RKKY interactions between spins are of the Heisenberg type, so the coefficients $J_x = J_y = J_z$ are all equal.
In triplet superconductors, the net spins of the Cooper pairs can lead to different RKKY interactions for different spin projections.
Since all states considered herein are invariant under rotations in the $xy$ plane, we can parametrize this as $\mathcal{F} = \bar{J}(\bm{S}_1\cdot\bm{S}_2) + \delta J (\bm{S}_1\cdot\bm{e}_z)(\bm{S}_2\cdot\bm{e}_z)$, where $\bar{J}$ is the Heisenberg part and $\delta J$ the Ising part.
Compared to the formulation $\mathcal{F} = \sum_n J_n S_{1n} S_{2n}$ of the free energy, this parametrization corresponds to $J_x = J_y = \bar{J}$ and $J_z = \bar{J} + \delta J$.

In \cref{fig:comparison}, we show the numerical results for $\bar{J}$ and $\delta J$ for each $p$-wave triplet order parameter listed in \cref{tab:orders}.
    For the $p_x$-wave state [\cref{fig:comparison}(a)], there is both a Heisenberg and Ising contribution.
    The Ising contribution dominates for most values of $\delta$, and has a strong bias towards positive (antiferromagnetic) values of $\delta J$.
    For the $p_y$-wave state [\cref{fig:comparison}(b)], the RKKY interaction is nearly purely Heisenberg-like, similarly to what one would observe in a normal metal.
    What is not so visible in the plot is that there exists a very weak Ising term $\delta J$ as well, which is consistently negative (ferromagnetic).
    This contribution is however two orders of magnitude smaller than the Heisenberg contribution, and it is therefore unclear to what extent it would be experimentally observable.
    For the chiral $p$-wave order [\cref{fig:comparison}(c)], the results are nearly identical to the $p_x$-wave order.
    This makes sense, because the order parameters of $p_x$-wave and chiral $p$-wave superconductors look identical for particles traveling along the $x$ axis.
    The core difference between these two states is that the chiral $p$-wave state would produce identical results also for spin displacements along the $y$ axis, in contrast to the $p_x$-wave state discussed above.

\begin{figure}[bt!]
    \includegraphics[width=\columnwidth]{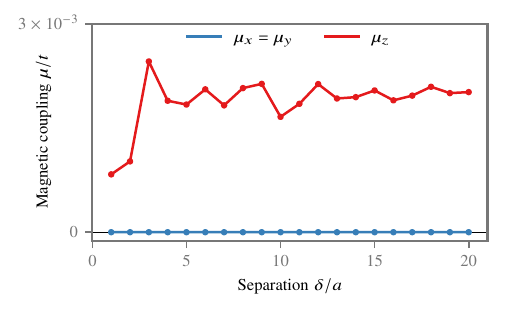}
    \caption{
        Magnetic interaction $\bm{\mu}$ between a non-unitary superconductor and each impurity.
        For this particular system, a magnetic coupling $\bm{\mu} \approx 0.002t\bm{e}_z$ was found which does not sensitively depend on the distance between the impurities.
        We generally found $\bm\mu = 0$ for unitary superconducting orders---including states that break time reversal symmetry, such as chiral $p$-wave and $d+is$-wave superconductors.
    }
    \label{fig:mu}
\end{figure}

Finally, we consider a non-unitary superconductor.
Non-unitary superconductors have two crucial differences from the other $p$-wave triplet states that we study~\cite{mackenzie2003a}.
Firstly, their Cooper pair condensate has a finite spin expectation value $\bm{\mu} \sim \bm{d}\times\bm{d}^*$, which can couple magnetically to each individual spin $\bm{S}_1$ and $\bm{S}_2$.
In \cref{fig:mu}, we show the numerical results for $\bm{\mu}$ obtained using the methodology developed in \cref{sec:numerics}.
Secondly, non-unitary superconductors generally have spin-dependent order parameters: $\Delta_{\up}(\bm{p}) \neq \Delta_{\dn}(\bm{p})$.
For the particular non-unitary state considered here, all relevant momenta $\bm{p} = p_x \bm{e}_x + p_y \bm{e}_y$ yield a full gap $|\Delta_{\up}| = \Delta_0$ in one spin band and no gap $|\Delta_{\dn}| = 0$ in the other.
In this case, both Cooper pairs and low-energy quasiparticles can always contribute to the RKKY interaction---regardless of what direction in the $xy$ plane the spins are separated along.
The consequence of this is visible in \cref{fig:comparison}(d): The Heisenberg interaction is very similar to in a normal metal due to the quasiparticle contribution, but there is also a significant Ising interaction due to the Cooper-pair contribution.
One might however ask why the Ising contribution appears to be negative (ferromagnetic), as opposed to e.g.\ the chiral $p$-wave case where a positive (antiferromagnetic) value was found.
This is partly a consequence of our parametrization.
When comparing the chiral state $\bm{d}(\bm{p}) = (p_x + ip_y) \bm{e}_z$ to the non-unitary state $\bm{d}(\bm{p}) = (1/2) (p_x + ip_y) (\bm{e}_x + \bm{e}_z)$, we might expect that a positive Ising term along the $z$ axis in the former case would correspond to positive Ising terms in along the $x$ and $y$ axes in the latter case.
Positive Ising terms along the $x$ and $y$ axes can equivalently be described as a negative Ising term along the $z$ axis, with a corresponding shift of the Heisenberg term~\footnote{To see this explicitly, we write out the free energy of a system with only Heisenberg and Ising interactions between two spins:
    \begin{align*}
      \mathcal{F} &= \bar{J} \bm{S}_1 \cdot \bm{S}_2 - \delta J \, S_{1z} S_{2z} \\
                  &= \bar{J} S_{1x} S_{2x} + \bar{J} S_{1y} S_{2y} + (\bar{J} - \delta J) S_{1z} S_{2z} \\
                  &= (\bar{J} - \delta J) \bm{S}_1 \cdot \bm{S}_2 + \delta J \, S_{1x} S_{2x} + \delta J \, S_{1y} S_{2y}.
    \end{align*}
    Thus, these two are equivalent: (i)~Heisenberg interaction $\bar{J}$ and  Ising interaction $-\delta J$ along the $z$ axis; (ii)~Heisenberg interaction $\bar{J}-\delta J$ and Ising interaction $+\delta J$ along the $x$ and $y$ axes.}.
This is precisely what we find in our numerical results.

\begin{figure}[bt!]
    \includegraphics[width=\columnwidth]{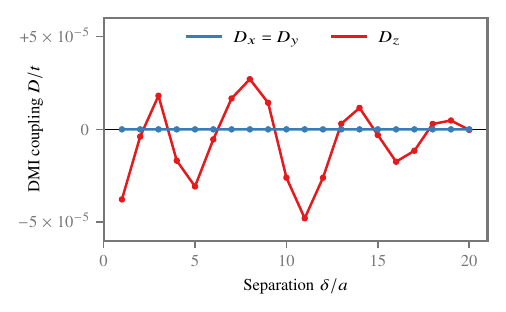}
    \caption{
        DMI-like interaction $\bm{D}\cdot(\bm{S}_1 \times \bm{S}_2)$ for the non-unitary superconductor.
        Similarly to the usual RKKY interaction, it oscillates as a function of the distance between the impurities.
        Similarly to the usual exchange-like RKKY interaction, we find that this new contribution oscillates between positive and negative values as a function of the distance between the two spins.
        We found no such interaction in unitary superconductors---including chiral $p$-wave and $d+is$-wave superconductors, which also break time reversal symmetry.
    }
    \label{fig:dmi}
\end{figure}

  On the other hand, the logic above only applies to the Ising contribution $\sim(\bm{d} \cdot \bm{S}_1)(\bm{d}^* \cdot \bm{S}_2)$ that arises in all $p$-wave triplet superconductors.
  The analytical results presented in \cref{sec:analytics} also uncovered a second Ising contribution $\sim (\bm{\mu} \cdot \bm{S}_1)(\bm{\mu} \cdot \bm{S}_2)$, where $\bm{\mu} \sim \bm{d} \times \bm{d}^*$ is the spin expectation value of the condensate.
  For the non-unitary superconductor discussed here this \emph{should} produce another Ising interaction along the $z$ axis.
  Thus, the non-unitary superconductor most likely has separate Ising interactions for spins aligned along the $z$ axis and spins aligned in the $xy$ plane---and these cannot be separated numerically in the presence of the background Heisenberg interaction.

\begin{figure}[bt!]
    \includegraphics[width=0.85\columnwidth]{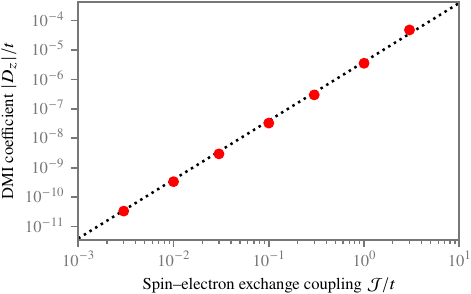}
    \caption{
            Magnitude of the Dzyaloshinskii--Moriya interaction $|D_z|$ as a function of the spin--electron exchange coupling~$\mathcal{J}$.
            These results were calculated for a distance $\delta = 11a$ between the two spins.
            All other model parameters are equal to \cref{fig:dmi}.
            Red dots are numerical results whereas the dotted line is a fit to a quadratic function $F(\mathcal J) = \mathcal{A} \mathcal J^2$.
    }
    \label{fig:logdmi}
\end{figure}

  The analytical calculations in \cref{sec:analytics,sec:specific} suggested that non-unitary superconductors may mediate effective Dzyaloshinskii--Moriya interactions $\sim \bm{D}\cdot(\bm{S}_1\times\bm{S}_2)$ between spins where $\bm{D} \sim \mathcal{J}^2 (\bm{d} \times \bm{d}^*) \sim \mathcal{J}^2 \bm{\mu}$.
  On the other hand, further analytical calculations revealed that $\bm{D} = 0$ for infinite translation-invariant superconductors.
  \Cref{fig:dmi} shows the numerically calculated DMI vector for the non-unitary superconductor, extracted using the methodology in \cref{sec:numerics}.
  Contrary to the analytics, these results show a \emph{finite} value for the DMI vector.
  Moreover, the DMI vector behaves exactly as one would expect from the analytical derivation: \cref{fig:dmi} shows that $\bm{D} \sim \bm{\mu} \sim \bm{e}_z$, whereas \cref{fig:logdmi} clearly shows that $\bm{D} \sim \mathcal{J}^2$.
  In contrast, no such DMI contribution was found for any of the unitary superconducting states that we studied herein.

\begin{figure}[bt!]
    \includegraphics[width=\columnwidth]{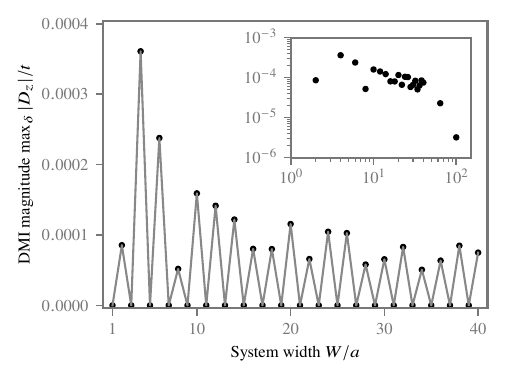}
    \caption{
          Magnitude of the DMI coefficient in a $100a\times W$ non-unitary superconductor as a function of the system width~$W$.
          The magnitude shown here is found by calculating the DMI vector $\bm{D}(\delta)$ for varying spin separation distances $\delta \in [1a, 20a]$ along the $x$ axis, and extracting the largest value $\max_\delta |D_z(\delta)|$ within that range.
          The inset replots the results as a log-log plot, and includes two extra points $W = 64a$ and $W=100a$ to show the trend as one approaches the 2D limit.
    }
    \label{fig:W}
\end{figure}

  To understand the precise origin of this DMI term, we ran a number of numerical simulations for systems of dimension $100a \times W$ for varying system widths~$W$.
  For comparison, the system studied above had dimensions $280a \times 40a$, corresponding to a relatively long and narrow superconductor.
  In \cref{fig:W}, we show the numerical results for the DMI coefficient as a function of the junction width.
  First, note that the DMI coefficient vanishes for $W=1a$ (1D limit) and $W \rightarrow \infty$ (2D limit).
  Only in the ``nanowire limit'' of long and narrow junctions does the DMI coefficient become sizeable.
  This is different from the parameter regime we studied analytically, where we assumed an infinitely large superconductor to make analytical progress.
  Next, we find a finite DMI coefficient \emph{only} for even values of~$W$, resulting in striking even--odd oscillations in \cref{fig:W}.
  This requires some more explanation.
  For a system of dimensions $100a \times W$, we choose to place the two spins at coordinates $\bm{R}_1 = (50a, \lfloor W/2 \rfloor)$ and $\bm{R}_2 = (50a + \delta, \lfloor W/2 \rfloor)$.
  Here, the coordinates are zero-indexed, meaning that the possible values along the $y$ axis are $\{ 0, 1, \ldots, W-2, W-1\}$.
  When $W$ is an odd number, the coordinates above imply that both spins are placed \emph{exactly} at the center of the superconductor along the $y$ axis, such that the system is mirror symmetric in that direction.
  In this case, following the symmetry arguments of Moriya (cf. rule 3 and 5 in Ref.~\cite{moriya1960a}), the DMI vector must be identically zero.
  On the other hand, when $W$ is an even number, there exists no $y$-coordinate on the lattice that corresponds to the exact center of the system, and we see that both spins are placed slightly off-center.
  In that case, a finite DMI vector is permitted, and we find numerically that a sizeable DMI coefficient appears.
  This is outside the regime of validity of our analytical calculation, where the assumption of an infinite superconductor with perfect translation symmetry precluded any such inversion symmetry breaking effects.
  This resolves the apparent contradiction between the numerical and analytical results regarding the DMI term in the RKKY interaction.

\begin{figure}[tb!]
    \includegraphics[width=\columnwidth]{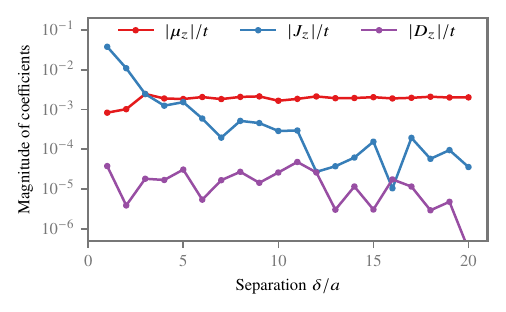}
    \caption{
        Magnitude of the magnetic interaction~$|\mu_z|$, Ising interaction~$|J_z|$, and Dzyaloshinskii--Moriya interaction~$|D_z|$ for the non-unitary superconductor (in units of $t$).
        For the parameters we have considered, the DMI contribution is found to be roughly two orders of magnitude lower than the other contributions to the free energy.
    }
    \label{fig:logmag}
\end{figure}

Let us now return to the DMI results for the $280a\times40a$ system shown in \cref{fig:dmi}.
Similarly to the usual RKKY interaction, this DMI contribution oscillates and decays as a function of the separation distance~$\delta$ between the two spins.
For this particular non-unitary state, the DMI prefers non-collinear spin orientations in the $xy$ plane.
In practice, the DMI vector obtained here is likely too small to be experimentally observable.
This is made clearer if one plots the magnitudes of the different terms in the free energy as a function of separation distance~$\delta$ (see \cref{fig:logmag}).
This shows that the DMI coefficient remains two orders of magnitude smaller than the other contributions to the free energy for all separation distances.
Whether the dominant contribution is an RKKY interaction or a magnetic interaction depends on the separation distance between the two spins.
Although the DMI interaction is too small to be observable for the parameter space we have explored, its existence motivates further studies of non-unitary superconductivity (either intrinsic or arising in hybrid structures) where it could be possible to find ways to increase the magnitude of the DMI interaction.

\begin{figure}[tb!]
    \includegraphics[width=\columnwidth]{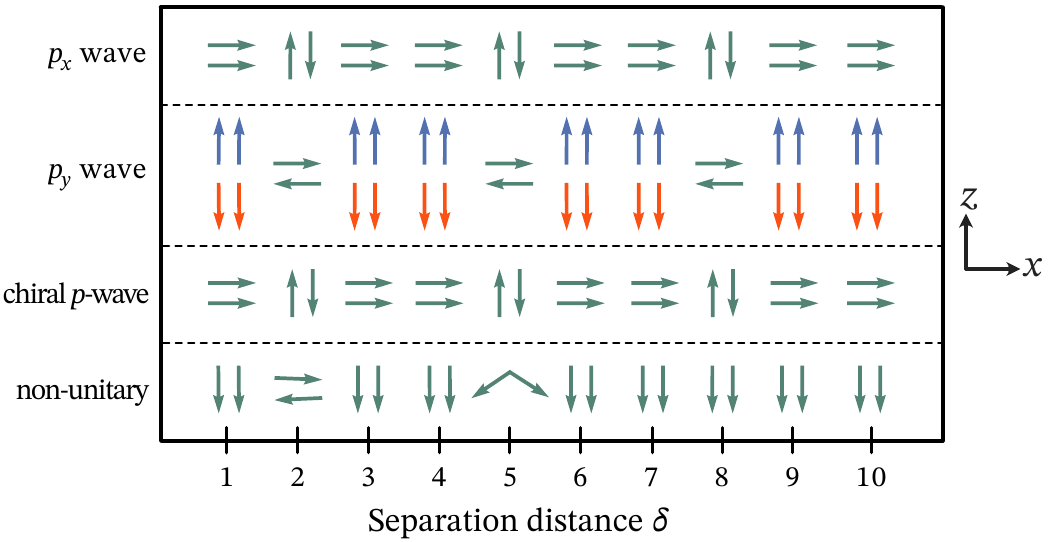}
    \caption{
        Ground-state orientation of the two spins $\bm{S}_1$ and $\bm{S}_2$ in \cref{fig:model} for the $p$-wave triplet order parameters in \cref{tab:orders}.
        For values of $\delta$ where a single ground state is found, we plot $\bm{S}_1$ and $\bm{S}_2$ as two green arrows.
        When two degenerate solutions exist, we plot one set of solutions as blue arrows and the other as orange arrows.
        We here focus on solutions in the $xz$ plane; the degenerate solutions that can be obtained by spin rotation in the $xy$ plane are not shown.
        Note that the numerical results indicate a negligible DMI coefficient $\bm{D}$ compared to the other terms in the free energy; in systems where this is not the case, also the $y$ components of the spins would be important.
    }
    \label{fig:groundstate}
\end{figure}

In \cref{fig:groundstate}, we plot the ground-state spin configuration for the various $d$-vectors presented in \cref{tab:orders}.
These configurations were obtained using the methodology presented in \cref{sec:groundstate}.
In normal metals and $s$-wave superconductors (not shown), the RKKY interaction oscillates between ferromagnetic and antiferromagnetic as a function of~$\delta$.
However, since it is a pure Heisenberg interaction $\sim\bm{S}_1 \cdot \bm{S}_2$, it is \emph{completely} degenerate with respect to simultaneous spin rotations of $\bm{S}_1$ and $\bm{S}_2$.
(Note however that we have neglected magnetic anisotropy herein, which could break this degeneracy in real systems.)
With a $p_y$-wave order parameter, the situation is \emph{nearly} the same as in a normal metal: The dominant contribution to the RKKY interaction is still the Heisenberg term.
However, we also find a very small ferromagnetic Ising term $\sim-(\bm{S}_1\cdot\bm{e}_z)(\bm{S}_2\cdot\bm{e}_z)$.
This term breaks the rotational symmetry in the $xz$ plane, and makes the system prefer ferromagnetic orientation along the $z$ axis but antiferromagnetic orientation along the $x$ axis.
Given the small magnitude of this Ising contribution for our parameter choices, this effect may not be visible in experiments.

Next, consider the $p_x$-wave order parameter.
The results are then exactly opposite from the $p_y$-wave case:
The system oscillates between ferromagnetic ordering along the $x$ axis and antiferromagnetic ordering along the $z$ axis.
However, the Ising interaction is in this case the same order of magnitude as the Heisenberg interaction (see \cref{fig:comparison}).
We therefore expect this oscillation between in-plane and out-of-plane spin ordering to be much more robust than for the $p_y$-wave order parameter, and to be experimentally observable.
The configurations we obtain for chiral $p$-wave and $p_x$ wave orders are identical---which we could expect as these materials look the same for quasiparticles propagating between $\bm{S}_1$ and $\bm{S}_2$ (i.e.\ along the $x$ axis).

Finally, let us discuss the results for the non-unitary superconductor.
One might expect behavior reminiscent of the $p_y$-wave superconductor since both of these materials have available zero-energy quasiparticle states.
Specifically, the non-unitary superconductor we considered has a gap in only one spin band, whereas the $p_y$ wave superconductor has no gap for propagation along the $x$ axis.
However, there are some important differences between these systems, which arise from the magnetization $\bm{\mu} \sim \bm{d}\times\bm{d}^*$ in the non-unitary superconductor.
Firstly, $\bm\mu$ breaks the symmetry between the $+z$ and $-z$ axes, so in the ferromagnetic state the spins always align along the $-z$ axis.
Secondly, this magnetization becomes the dominant energy scale for long separation distances $\delta$, and thus the oscillation between ferromagnetic and antiferromagnetic states stops {beyond some distance}.
Finally, when the spins prefer antiferromagnetic alignment along the $x$ axis, the $\mu_z(S_{1z} + S_{2z})$ interaction yields an additional ferromagnetic coupling along the $z$ axis.
This produces non-collinearity.
Specifically, for $\delta=2a$ and $\delta=5a$, respectively, the precise angle between the two spins in the ground state is $175.3^\circ$ and $113.1^\circ$.
If the DMI vector had been larger, an additional non-collinearity could have been found in the $xy$ plane.
However, in our particular simulation results, this effect was found to be negligible.

The RKKY interaction can be measured experimentally following the method of Ref. \cite{zhou_natphys_10}. 
Consider two spin impurities placed on a surface.
The electron tunneling current from one of these impurities to a spin-polarized tip depends on the impurity spin direction.
If the tip and impurity have the same spin direction, the tunneling current will be large since the available density of states is large.
Conversely, if they have the opposite spin direction, the tunneling current will be small.
By measuring the tunneling current from each impurity to the tip, in the presence of an external magnetic field, one then expects a different $\mathrm{d}I/\mathrm{d}V$ curve depending on whether the impurity spins are aligned parallell (P) or antiparallell (AP).
The measured $\mathrm{d}I/\mathrm{d}V$ curve (in particular its central peak) can subsequently be directly related to the spin orientation of the spin impurity, allowing one to extract how the spin direction changes in the presence of an applied external field.
If the spins are AP in the absence of field, their AP coupling will only be broken when a sufficiently strong external field $B=B_c$ is applied.
The AP coupling therefore modifies the $\mathrm{d}I/\mathrm{d}V$ curve at small field values $|B|<|B_c|$.
This $\mathrm{d}I/\mathrm{d}V$ curve is then used to extract a magnetization curve, i.e. the spin as a function of the applied magnetic field.
The effect above results in opposite signs for the slope of the magnetization curve for small and high values of $|B|$, respectively, and can even cause the magnetization to change sign for small $|B| < |B_c|$ compared to $|B| > |B_c|$.
This is precisely because the tunneling current is now smaller between this impurity and the tip.
The $\mathrm{d}I/\mathrm{d}V$ curve for the other impurity gives rise to a magnetization curve which has the standard paramagnetic behavior corresponding to the P alignment, since the other impurity is aligned with the tip.

Now, the critical field $B_c$ where the magnetization curve vs. $B$ (extracted from the measured $\mathrm{d}I/\mathrm{d}V$) changes its qualitative characteristics in the AP configuration is directly related to the RKKY energy $E_\text{RKKY}$ of the impurity spins. When $E_\text{RKKY}$ equals the Zeeman-energy induced by $B_c$, the change occurs. In this way, by measuring the $\mathrm{d}I/\mathrm{d}V$ curves in the presence of a magnetic field, one can extract $E_\text{RKKY}$ in the AP configuration. In the P configuration, the exchange coupling $\mathcal{J}$ between the impurity spins and conduction electrons must instead be inferred by fitting the $\mathrm{d}I/\mathrm{d}V$ data to a theoretical model.

\section{Conclusion}
We have studied the RKKY interactions in $p$-wave triplet superconductors numerically and analytically.
Similarly to singlet superconductors, we find that the momentum dependence of the superconducting gap $|\Delta(\bm{p})|$ strongly affects the RKKY interaction.
However, in triplet superconductors the order parameter also develops a spin structure.
We have shown that this generally leads to large Ising-like RKKY interactions between spins placed on the surface of the superconductor.
In addition, we find that a DMI-like RKKY interaction arises for non-unitary triplet superconductors when the impurity spins are placed sufficiently close to an edge of the system (which breaks inversion symmetry).

An interesting venue for further research would be to determine whether the DMI contribution could be enhanced in other physical setups, as the effect of the DMI vector on the magnetic ground state was found to be negligible for the parameters studied herein.
Despite a negligible DMI effect, we find that a non-collinear spin configuration arises in non-unitary triplet superconductors due to a combination of antiferromagnetic in-plane RKKY interactions and ferromagnetic out-of-plane interactions with the Cooper pair condensate.
Our results suggest that the RKKY interaction in $p$-wave triplet superconductors can be used as probe a probe for both the spin and momentum symmetries of the superconducting order parameter.

\begin{acknowledgments}
  This work was supported by the Research Council of Norway through Grant No.\ 323766 and its Centres of Excellence funding scheme Grant No.\ 262633 ``QuSpin.''
  {This work was also supported by JSPS KAKENHI Grant Number JP30578216 and the JSPS-EPSRC Core-to-Core program ``Oxide Superspin".}
  The numerical calculations were performed on resources provided by Sigma2---the National Infrastructure for High Performance Computing and Data Storage in Norway, project NN9577K.
\end{acknowledgments}

\appendix
\section{Efficient calculation of the LDOS}\label{sec:ldos}
We now describe the calculation procedure for the  local density of states (LDOS).
One way to define the Green function is as the \emph{resolvent} of the Hamiltonian~\cite{wu1993a,nagai2017a,covaci2010a}.
Given the $4N\times4N$ Hamiltonian matrix~$\check{H}$ introduced in \cref{sec:tight-binding}, the Green function matrix on the lattice is then given by
\begin{equation}
    \label{eq:resolvent}
    [(\omega+i\eta)\check{I} - \check{H}] \check{G}^R(\omega) = \check{I},
\end{equation}
where $\check{I}$ is an identity matrix, $\omega$ is the energy, and $\eta \rightarrow 0^+$ yields the retarded Green function.
In practice, we let $\eta$ go to a small but finite value; this broadens the $\delta$-function peaks in the LDOS to finite-width Lorentzians, which are easier to evaluate numerically.
From the diagonal elements of this $4N\times4N$ Green function matrix, we can then obtain the spin-resolved LDOS $N_{i\sigma}$ at every lattice site~$i$~\cite{nagai2017a,covaci2010a}:
\begin{align}
    N_{i\up}(+\omega) &= -\frac{1}{\pi}\mathrm{Im}[G^R_{4i+1,4i+1}(\omega)], \\
    N_{i\dn}(+\omega) &= -\frac{1}{\pi}\mathrm{Im}[G^R_{4i+2,4i+2}(\omega)], \\
    N_{i\up}(-\omega) &= -\frac{1}{\pi}\mathrm{Im}[G^R_{4i+3,4i+3}(\omega)], \\
    N_{i\dn}(-\omega) &= -\frac{1}{\pi}\mathrm{Im}[G^R_{4i+4,4i+4}(\omega)],
\end{align}
Since the Green function contains information about both electrons and holes when we use the Nambu-space formulation, we only need to determine the Green function for $\omega \geq 0$ to plot the LDOS for all~$\omega$.
Moreover, to study YSR states, calculating the LDOS for $\omega \in [0, \Delta_0]$ is usually sufficient.

The most straight-forward way to solve \cref{eq:resolvent} for the Green function is direct matrix inversion: $\check{G}^R(\omega)  = [(\omega+i\eta)\check{I} - \check{H}]^{-1}$.
In practice, this is prohibitively expensive {numerics} for large systems.
Instead, let us exploit the property that we only require the Green function elements $G^R_{n,n}$ for a few specific indices $n$ in order to determine the LDOS at one specific lattice site~$i$.
If we split the Green function matrix $\check{G}^R = [\check{x}_1 \cdots \check{x}_{4N}]$ into column vectors and the identity matrix $\check{I} = [\check{e}_1 \cdots \check{e}_{4N}]$ into unit vectors, then \cref{eq:resolvent} can equivalently be written
\begin{equation}
    \check{A}(\omega) \, \check{x}_n(\omega) = \check{e}_n,
\end{equation}
where $\check{A}(\omega) \equiv (\omega+i\eta)\check{I} - \check{H}$.
This is a standard linear system of equations, which can be solved very efficiently using e.g.\ the SciPy~\cite{virtanen2020a} function \texttt{scipy.sparse.linalg.spsolve} if $\check{A}$ and $\check{e}_n$ are stored as sparse matrices.
Once these equations have been solved for $n \in \{4i+1, \ldots, 4i+4\}$, the LDOS at lattice site~$i$ is easily found from the result vectors:
\begin{align}
    N_{i\up}(+\omega) &= -\frac{1}{\pi}\mathrm{Im}[\check{e}_{4i+1}\cdot\check{x}_{4i+1}(\omega)], \\
    N_{i\dn}(+\omega) &= -\frac{1}{\pi}\mathrm{Im}[\check{e}_{4i+2}\cdot\check{x}_{4i+2}(\omega)], \\
    N_{i\up}(-\omega) &= -\frac{1}{\pi}\mathrm{Im}[\check{e}_{4i+3}\cdot\check{x}_{4i+3}(\omega)], \\
    N_{i\dn}(-\omega) &= -\frac{1}{\pi}\mathrm{Im}[\check{e}_{4i+4}\cdot\check{x}_{4i+4}(\omega)].
\end{align}
This approach provides an $\mathcal{O}(N)$ computational speedup over a full matrix inversion $\check{G}^R(\omega) = \check{A}^{-1}(\omega)$, since we now only calculate the specific columns of $\check{G}^R(\omega)$ that are required to evaluate the \emph{local} density of states at one or a few sites.
For moderately large lattices ($10^4$--$10^5$ sites), we found that this approach outperforms direct matrix inversion (as discussed above), matrix diagonalization (the conventional approach), and Chebyshev matrix expansion \cite{weisse2006a,covaci2010a}.

The approach above is inspired by Nagai et al.~\cite{nagai2017a}.
Their algorithm is likely more efficient, as it re-uses results between different $\omega$ values, whereas in our case the calculation at each $\omega$ is independent.
The procedure we used is however much simpler, as it can be implemented in few lines of code that leverages standard numerical libraries.
This approach was found to be sufficiently fast for our purposes:
Calculating the LDOS for 100 energies at one site of a $200\times200$ lattice requires $\sim\! 2$~min computation time on a modern desktop computer.
While smaller $64\times64$ lattice sizes were used for the initial parts of the parameter optimization described in \cref{sec:params}, the final plot presented in \cref{fig:ysr} corresponds to a much larger $800\times800$ lattice size.
For comparison, the total computation time to generate \cref{fig:ysr} was $\sim\! 4000$ CPU hours, which was still quite feasible as an overnight simulation on an HPC cluster.

\section{Derivation of the RKKY interaction}\label{sec:general}
We here provide a detailed derivation of the RKKY interaction energy $E_{\text{RKKY}}$ based on Green function methods.
At a high level, our approach is similar to previous derivations by e.g.\ Schwabe et al.~\cite{schwabe1996a}.
However, in contrast to Ref.~\cite{schwabe1996a}, our derivation uses the Keldysh formalism~\cite{rammer1986a} and includes more intermediate steps to make the derivation more accessible.

Our starting point is the following argument.
Consider a uniform {superconductor} described by an unperturbed Green function~$\check{G}_0$.
We now place a classical spin $\bm{S}_1$ at a position $\bm{R}_1$, which couples to the electrons in the metal via an exchange interaction $\mathcal{H}_{\text{int}} = -(\mathcal{J}/2) \bm{S}_1 \cdot \bm{s}(\bm{R}_1)$, where $\bm{s}(\bm{r})$ is the electron spin operator at a position~$\bm{r}$.
This perturbs the Green function from $\check{G}_0$ to $\check{G} \equiv \check{G}_0 + \delta\check{G}$.
This $\delta\check{G}$ then shifts the electron spin density by an amount $\delta\bm{s}_1(\bm{r})$, which typically oscillates and decays as a function of the distance $|\bm{r} - \bm{R}_1|$ from the spin~$\bm{S}_1$.
If we then place a second spin $\bm{S}_2$ at $\bm{r} = \bm{R}_2$, this spin interacts with the electron spin density $\delta\bm{s}_1(\bm{R}_2)$ at that position via a second exchange interaction $-(\mathcal{J}/2) \bm{S}_2 \cdot \delta\bm{s}_1(\bm{R}_2)$.
Assuming that $\delta\bm{s}_1$ has been calculated to leading order $\mathcal{O}(\mathcal{J})$, the result is an $\mathcal{O}(\mathcal{J}^2)$ contribution to $E_{\text{RKKY}}$.
When we also include the inverse process, i.e.\ how the perturbation $\delta\bm{s}_2$ generated by $\bm{S}_2$ affects $\bm{S}_1$, the RKKY interaction energy can be written:
\begin{equation}
    \label{eq:rkky}
    \begin{aligned}
        E_{\text{RKKY}} & = E_{\text{RKKY}}^{21} + E_{\text{RKKY}}^{12}                                                          \\
                        & = (-\mathcal{J}/2)[\bm{S}_1 \cdot \delta\bm{s}_2(\bm{R}_1) + \bm{S}_2 \cdot \delta\bm{s}_1(\bm{R}_2)].
    \end{aligned}
\end{equation}
Below, we only calculate $E_{\text{RKKY}}^{12} \sim \bm{S}_2 \cdot \delta\bm{s}_1(\bm{R}_2)$ explicitly, as the remaining term $E_{\text{RKKY}}^{21}$ follows from symmetry.

Note that some terms of similar order have been discarded in the argument above, as we are only interested in deriving the interactions between $\bm{S}_1$ and $\bm{S}_2$.
First, in ferromagnets and non-unitary superconductors, even the unperturbed Green function $\check{G}_0$ gives rise to a finite spin expectation value $\bm{s}(\bm{r})$.
This yields a lower-order contribution $(-\mathcal{J}/2)[\bm{S}_1 \cdot \bm{s}(\bm{R}_1) + \bm{S}_2 \cdot \bm{s}(\bm{R}_2) ]$ to the energy of the system, which is important for understanding the ground-state spin configuration.
Second, even in non-magnetic metals, there are two more contributions of the same order as we consider: $(-\mathcal{J}/2)[\bm{S}_1 \cdot \delta\bm{s}_1(\bm{R}_1) + \bm{S}_2 \cdot \delta\bm{s}_2(\bm{R}_2)]$.
These terms can be understood as a paramagnetic interaction between the metal and each individual spin.
However, none of these contributions correspond to interactions between the two spins $\bm{S}_1$ and $\bm{S}_2$, and are not considered RKKY interactions.

\subsection{Green function formalism}
As mentioned above, we here employ the Keldysh Green function formalism~\cite{rammer1986a}---just generalized from Keldysh$\otimes$Spin space to Keldysh$\otimes$Nambu$\otimes$Spin space, as required to describe superconductors.
We thus define the $8\times8$ matrices
\begin{align}
    \check{G}                    & =
    \begin{pmatrix}
        \hat{G}^R & \hat{G}^K \\
                  & \hat{G}^A
    \end{pmatrix},     &
    \check{G}_0                  & =
    \begin{pmatrix}
        \hat{G}_0^R & \hat{G}_0^K \\
                    & \hat{G}_0^A
    \end{pmatrix}, &
    \check{\Sigma}               & =
    \begin{pmatrix}
        \hat{\Sigma}^R & \hat{\Sigma}^K \\
                       & \hat{\Sigma}^A
    \end{pmatrix},
\end{align}
where quantities with hats are $4\times4$ matrices in Nambu$\otimes$Spin space.
We now expand the Dyson equation for $\check{G}$ to first order in the self-energy~$\check{\Sigma}$~\cite{rammer1986a}, which yields the following shift $\delta\check{G} = \check{G} - \check{G}_0$ from the unperturbed Green function~$\check{G}_0$:
\begin{equation}
    \delta\check{G} = \check{G}_0 \otimes \check{\Sigma} \otimes \check{G}_0.
\end{equation}

Next, consider a classical spin $\bm{S}_1$ placed at~$\bm{R}_1$ on the {superconductor}.
This can be modeled using the block-diagonal self-energy matrix $\check{\Sigma} = \mathrm{diag}(\hat{\Sigma}, \hat{\Sigma})$, where the $4\times4$ self-energy
\begin{equation}
    \label{eq:self-energy}
    \hat{\Sigma}(1, 2) = - (\mathcal{J} / 2) (\bm{S}_1 \cdot \hat{\bm{\sigma}})\, \delta(\bm{r}_1 - \bm{R}_1)\, \delta(1 - 2).
\end{equation}
We here use the common short-hand notation ${1 \rightarrow (\bm{r}_1, t_1)}$ and ${2 \rightarrow (\bm{r}_{2}, t_{2})}$ for space-time coordinates~\cite{rammer1986a}.
Moreover, we have introduced a vector of $4\times4$ matrices $\hat{\bm{\sigma}} = \mathrm{diag}(\bm{\sigma}, \bm{\sigma}^*)$~\footnote{We define the Green functions such that the Gorkov equation takes the form $[i\hat{\tau}_3 \partial_t - \hat{H}]\check{G}(1,2) = {\delta(1-2)}$, where $\hat{\tau}_3$ is the third Pauli matrix in Nambu space.
Some authors do not include~$\hat{\tau}_3$ here, in which case the sign structure of $\check{G}$ changes.
One must then compensate by including an additional sign difference between electrons and holes such that $\hat{\bm{\sigma}} \rightarrow \mathrm{diag}(+\bm{\sigma}, -\bm{\sigma}^*)$.}, where $\bm\sigma$ is the Pauli vector.
It can be shown that this self-energy matrix~$\check{\Sigma}$ correctly reproduces the Gorkov equation for a metal with a spin impurity~\cite{gorkov1958a,morten2003a}.
The definition of ``$\otimes$''~\cite{rammer1986a} now yields the more explicit equation
\begin{equation}
    \delta\check{G}(1, 2) = \int d1' \int d2'\, \check{G}_0(1, 1') \, \check{\Sigma}(1', 2') \, \check{G}_0(2', 2),
\end{equation}
where $d1' \equiv d\bm{r}_{1'} \, dt_{1'}$ and so on.
For our purposes, we do not need to determine the whole $8\times8$ matrix $\delta\check{G}$; it is sufficient to calculate the top-right $4\times4$ block $\delta\hat{G}^K(1,2)$.
From direct multiplication of the definitions of these matrices, we find
\begin{equation}
    \begin{aligned}
        \delta\hat{G}^K(1, 2) =
        \int d1' \int d2'\, \Big\{
            & \hat{G}_0^R(1, 1') \, \hat{\Sigma}(1', 2') \, \hat{G}_0^K(2', 2) \\
        +\, & \hat{G}_0^K(1, 1') \, \hat{\Sigma}(1', 2') \, \hat{G}_0^A(2', 2)
        \Big\},
    \end{aligned}
\end{equation}
where $\hat{\Sigma}$ is given by \cref{eq:self-energy}.
Since $\hat{\Sigma}(1', 2') \sim {\delta(1'-2')}$, the integral over $2'$ is trivial.
Moreover, we only require the equal-coordinate Green function $\delta \hat{G}^K(1) \equiv \lim_{2 \rightarrow 1} \delta\hat{G}^K(1, 2)$ to determine the spin density $\delta \bm{s}_1$ that arises.
This is because the Green function $\hat{G}^K(1,2)$ is formally defined in terms of expectation values on the form $\langle \psi^\dagger_{\sigma_1}(\bm{r}_1, t_1)\, \psi^{\vphantom{\dagger}}_{\sigma_2}(\bm{r}_2, t_2) \rangle$, which only correspond to spin-resolved electron number operators when $(\bm{r}_1, t_1, \sigma_1)$ and $(\bm{r}_2, t_2, \sigma_2)$ are equal.
After some relabeling of the remaining coordinates, we thus obtain the equation
\begin{equation}
    \label{eq:delta-keldysh}
    \begin{aligned}
        \delta \hat{G}^K(1) =
        -\frac{\mathcal{J}}{2} \lim_{3 \rightarrow 1} \int & dt_2 \int d\bm{r}_2\,\delta(\bm{r}_2 - \bm{R}_1)                             \\
        \times \,
        \Big\{\;
                                                           & \hat{G}_0^R(1, 2) \, (\bm{S}_1 \cdot \hat{\bm{\sigma}}) \, \hat{G}_0^K(2, 3) \\
        +\,                                                & \hat{G}_0^K(1, 2) \, (\bm{S}_1 \cdot \hat{\bm{\sigma}}) \, \hat{G}_0^A(2, 3)
        \;\Big\}.
    \end{aligned}
\end{equation}

\subsection{Wigner transformation}
The next step is to perform a \emph{Wigner transformation}~\cite{kita2010a}.
To keep the notation simple, let us consider the following equation for now, which captures the overall mathematical structure of the actual equation for $\delta\hat{G}^K$ presented in \cref{eq:delta-keldysh}:
\begin{equation}
    \begin{aligned}
        A(\bm{r}_1, t_1)
        = \lim_{3 \rightarrow 1}
         & \int dt_2
        \int d\bm{r}_2\,\delta(\bm{r}_2 - \bm{R}_1)\, \\
         & \times B(\bm{r}_1, t_1 | \bm{r}_2, t_2) \,
        (\bm{S}_1 \cdot \hat{\bm{\sigma}})\,
        C(\bm{r}_2, t_2 | \bm{r}_3, t_3).
    \end{aligned}
\end{equation}
We now introduce the following center-of-mass coordinates ($\bm{r}_{ij}, t_{ij}$) and relative coordinates ($\bm{\rho}_{ij}, \tau_{ij}$):
\begin{equation}
    \begin{aligned}
        \bm{r}_{ij}    & \equiv (\bm{r}_i + \bm{r}_j)/2, &
        \bm{\rho}_{ij} & \equiv \bm{r}_i - \bm{r}_j,       \\
        t_{ij}         & \equiv (t_i + t_j)/2,           &
        \tau_{ij}      & \equiv t_i - t_j.                 \\
    \end{aligned}
\end{equation}
In terms of these so-called \emph{mixed coordinates}, we now have
\begin{equation}
    \begin{aligned}
        A(\bm{r}_1, t_1) & =                                                           \\
        \lim_{3 \rightarrow 1}
                         & \int dt_2
        \int d\bm{r}_2\,\delta(\bm{r}_2 - \bm{R}_1)\,                                  \\
                         & \times B(\bm{r}_{12}, \bm{\rho}_{12}, t_{12}, \tau_{12}) \,
        (\bm{S}_1 \cdot \hat{\bm{\sigma}})\,
        C(\bm{r}_{23}, \bm{\rho}_{23}, t_{23}, \tau_{23}).
    \end{aligned}
\end{equation}
Next, we Fourier transform relative coordinates $(\bm{\rho}_{ij}, \tau_{ij})$ into corresponding momentum and energy variables $(\bm{p}_{ij}, \omega_{ij})$,
\begin{equation}
    \begin{aligned}
        A(\bm{r}_1, t_1)       & =                                                         \\
        \lim_{3 \rightarrow 1} & \int dt_2 \int d\bm{r}_2\,\delta(\bm{r}_2 - \bm{R}_1)\,   \\
        \times                 & \int d\omega_{12} \, e^{i\omega_{12} \tau_{12}}
        \int d\omega_{23} \, e^{i\omega_{23} \tau_{23}}                                    \\
        \times                 & \int d\bm{p}_{12} \, e^{-i\bm{p}_{12}\cdot\bm{\rho}_{12}}
        \int d\bm{p}_{23} \, e^{-i\bm{p}_{23}\cdot\bm{\rho}_{23}} \,                       \\
        \times                 & \; B(\bm{r}_{12}, \bm{p}_{12}, t_{12}, \omega_{12}) \,
        (\bm{S}_1 \cdot \hat{\bm{\sigma}})\,
        C(\bm{r}_{23}, \bm{p}_{23}, t_{23}, \omega_{23}).
    \end{aligned}
\end{equation}
Let us now take the limit $\lim_{3 \rightarrow 1} = \lim_{\,\bm{r}_3 \rightarrow \bm{r}_1} \lim_{t_3 \rightarrow t_1}$.
The implications for the relative variables are that
\begin{equation}
\begin{aligned}
  \tau_{23} &\equiv t_2 - t_3 &\rightarrow&& t_2 - t_1 &\equiv -\tau_{12}, \\
  \bm{\rho}_{23} &\equiv \bm{r}_2 - \bm{r}_3 &\rightarrow&& \bm{r}_2 - \bm{r}_1 &\equiv -\bm{\rho}_{12}.
\end{aligned}
\end{equation}
Moreover, the actual functions $B$ and $C$ we will consider later correspond to components of the \emph{unperturbed} Green function $\check{G}_0$ of the system---which in our case is assumed to be homogeneous and time independent.
This means that $B$ and $C$ are in practice not functions of $\bm{r}_{12}, \bm{r}_{23}, t_{12}, t_{23}$.
With these considerations in mind, the result above simplifies to
\begin{equation}
\begin{aligned}
  A(\bm{r}_1,& t_1) = \\
    & \int dt_2 \int d\bm{r}_2 \, \delta(\bm{r}_2 - \bm{R}_1) \\
  \times & \int d\omega_{12} \int d\omega_{23} \, e^{i \tau_{12} (\omega_{12} - \omega_{23})} \\
  \times & \int d\bm{p}_{12} \int d\bm{p}_{23} \, e^{-i \bm{\rho}_{12} \cdot (\bm{p}_{12} - \bm{p}_{23})} \\
  \times & \; B(\bm{p}_{12}, \omega_{12}) \,
      (\bm{S}_1 \cdot \hat{\bm{\sigma}})\,
      C(\bm{p}_{23}, \omega_{23}).
\end{aligned}
\end{equation}
Next, we consider the integrals over $t_2$ and $\bm{r}_2$.
Since $t_1$ and $\bm{r}_1$ are held constant during the integration on the right-hand side of the equation, we can write $dt_2 = d(t_2 - t_1) \equiv -d\tau_{12}$ and $d\bm{r}_2 = d(\bm{r}_2 - \bm{r}_1) \equiv -d\bm{\rho}_{12}$.
Thus, we can rewrite the above as
\begin{equation}
\begin{aligned}
  A(\bm{r}_1,& t_1) = \\
    & \int d\tau_{12} \int d\bm{\rho}_{12} \, \delta(-\bm{\rho}_{12} + \bm{r}_1 - \bm{R}_1) \\
  \times & \int d\omega_{12} \int d\omega_{23} \, e^{i \tau_{12} (\omega_{12} - \omega_{23})} \\
  \times & \int d\bm{p}_{12} \int d\bm{p}_{23} \, e^{-i \bm{\rho}_{12} \cdot (\bm{p}_{12} - \bm{p}_{23})} \\
  \times & \; B(\bm{p}_{12}, \omega_{12}) \,
      (\bm{S}_1 \cdot \hat{\bm{\sigma}})\,
      C(\bm{p}_{23}, \omega_{23}).
\end{aligned}
\end{equation}
Note that the integrand only depends on $\tau_{12}$ via the complex exponential.
Using the Fourier identity for the delta function,
\begin{equation}
\int d\tau_{12} \, e^{i\tau_{12} (\omega_{12} - \omega_{23})} = 2\pi \, \delta(\omega_{12} - \omega_{23}),
\end{equation}
and subsequently integrating out $\omega_{23}$, we get
\begin{equation}
\begin{aligned}
A(\bm{r}_1,& t_1) = \\
2\pi & \int d\bm{\rho}_{12} \, \delta(-\bm{\rho}_{12} + \bm{r}_1 - \bm{R}_1) \\
\times & \int d\omega_{12} \int d\bm{p}_{12} \int d\bm{p}_{23} \, e^{-i \bm{\rho}_{12} \cdot (\bm{p}_{12} - \bm{p}_{23})} \\
\times & \; B(\bm{p}_{12}, \omega_{12}) \,
  (\bm{S}_1 \cdot \hat{\bm{\sigma}})\,
  C(\bm{p}_{23}, \omega_{12}).
\end{aligned}
\end{equation}
Next, we integrate out the spatial variable $\bm{\rho}_{12}$.
The remaining delta function ensures that $\bm{\rho}_{12} \rightarrow \bm{r}_1 - \bm{R}_1$, so we obtain
\begin{equation}
 \begin{aligned}
   A(\bm{r}_1,& t_1) = \\
   2\pi & \int d\omega_{12} \int d\bm{p}_{12} \int d\bm{p}_{23} \, e^{-i (\bm{r}_1 - \bm{R}_1) \cdot (\bm{p}_{12} - \bm{p}_{23})} \\
   \times & \; B(\bm{p}_{12}, \omega_{12}) \,
       (\bm{S}_1 \cdot \hat{\bm{\sigma}})\,
       C(\bm{p}_{23}, \omega_{12}).
 \end{aligned}
\end{equation}
For simplicity, we at this point rename the integration variables $\omega_{12} \rightarrow \omega, \bm{p}_{12} \rightarrow \bm{p}_1, \bm{p}_{23} \rightarrow \bm{p}_2$.
As there is no dependence on the absolute time~$t_1$, this can also be omitted on the left-hand side.
The final form of the Wigner transformation is thus
\begin{equation}
 \begin{aligned}
   A(\bm{r}_1) =
   2\pi \int d\omega \int & d\bm{p}_{1} \int d\bm{p}_{2} \, e^{i (\bm{p}_2 - \bm{p}_1) \cdot (\bm{r}_1 - \bm{R}_1)} \\
   \times & \; B(\bm{p}_{1}, \omega) \,
       (\bm{S}_1 \cdot \hat{\bm{\sigma}})\,
       C(\bm{p}_{2}, \omega).
 \end{aligned}
\end{equation}
We now apply this transformation to the Green function shift that we obtained in \cref{eq:delta-keldysh}, which yields the result
\begin{equation}
    \label{eq:delta-keldysh-result}
    \begin{aligned}
        \delta\hat{G}^K(\bm{r}_1) = - {\mathcal{J}\pi}
        \int & d\omega \int d\bm{p}_{1} \int d\bm{p}_{2} \,
        e^{{i}(\bm{p}_{2} - \bm{p}_{1})\cdot(\bm{r}_1 - \bm{R}_1)} \\
        \times\,
        \Big\{\,
             & \hat{G}_0^R(\bm{p}_1, \omega)
        \, (\bm{S}_1 \cdot \hat{\bm{\sigma}})
        \, \hat{G}_0^K(\bm{p}_2, \omega)
        \\ +\,
             & \hat{G}_0^K(\bm{p}_1, \omega)
        \, (\bm{S}_1 \cdot \hat{\bm{\sigma}})
        \, \hat{G}_0^A(\bm{p}_2, \omega)
        \,\Big\}.
    \end{aligned}
\end{equation}

\subsection{Spin expectation value}
Once the equal-coordinate Keldysh Green function $\hat{G}^K(\bm{r})$ is known, the electron spin density follows directly from~\footnote{See e.g.\ the supplemental of Ref.~\cite{ouassou2019e} for a complete derivation.
    Note that we here formulate the equation in terms of the $4\times4$ matrix $\hat{G}^K$ in Spin$\otimes$Nambu space, not its top-left $2\times2$ block~$G^K$.}
\begin{equation}
    \bm{s}(\bm{r}) = \frac{1}{8} \text{Im Tr}[\hat{\bm{\sigma}} \hat{G}^K(\bm{r})].
\end{equation}
Thus, in the kind of system treated in the previous subsections, we find the following perturbation of the spin density
\begin{equation}
    \delta \bm{s}_1(\bm{r}) = \frac{1}{8} \text{Im Tr}[\hat{\bm{\sigma}} \, \delta \hat{G}^K(\bm{r})].
\end{equation}
The contribution to \cref{eq:rkky} then becomes
\begin{equation}
    \begin{aligned}
        E_{\text{RKKY}}^{12}
         & = -\frac{\mathcal{J}}{2} \bm{S}_2 \cdot \delta \bm{s}_1(\bm{R}_2)                                         \\
         & = -\frac{\mathcal{J}}{16} \text{Im Tr}[(\bm{S}_2 \cdot \hat{\bm{\sigma}}) \, \delta \hat{G}^K(\bm{R}_2)].
    \end{aligned}
\end{equation}
We can now substitute \cref{eq:delta-keldysh-result} into this result, and find
\begin{equation}
    \begin{aligned}
        E_{\text{RKKY}}^{12} = \phantom{-J} &                                                                    \\
        \frac{{\pi}\mathcal{J}^2}{{16}} \, \mathrm{Im}
        \int                                & d\omega \int d\bm{p}_{1} \int d\bm{p}_{2} \,
        e^{{i}(\bm{p}_{2} - \bm{p}_{1})\cdot(\bm{R}_2 - \bm{R}_1)}                                                \\
        \times\, \mathrm{Tr}\,
        \Big\{\,
                                            & (\bm{S}_2\cdot\hat{\bm{\sigma}})\, \hat{G}_0^R(\bm{p}_1, \omega)
        \, (\bm{S}_1 \cdot \hat{\bm{\sigma}})
        \, \hat{G}_0^K(\bm{p}_2, \omega)
        \\ +\,
                                            & (\bm{S}_2 \cdot \hat{\bm{\sigma}})\, \hat{G}_0^K(\bm{p}_1, \omega)
        \, (\bm{S}_1 \cdot \hat{\bm{\sigma}})
        \, \hat{G}_0^A(\bm{p}_2, \omega)
        \,\Big\}.
    \end{aligned}
\end{equation}
Note that this is a quite general result:
We have performed a leading-order perturbation expansion in $\mathcal{J}$, but have not made any assumptions about the unperturbed system described by~$\check{G}_0$ except for homogeneity in space and time.
Thus, the same equation can be used to describe e.g.\ superconductors, magnets, or spin--orbit-coupled systems.
Moreover, the result obtained at this point is valid both in and out of equilibrium.

\subsection{Thermodynamic equilibrium}
The Keldysh Green function can be parametrized in terms of a $4\times4$ distribution function matrix~$\hat{h}$~\cite{rammer1986a,belzig1999a},
\begin{equation}
    \hat{G}_0^K = \hat{G}^R_0 \, \hat{h} - \hat{h} \, \hat{G}_0^A.
\end{equation}
In equilibrium, electrons follow the Fermi--Dirac distribution, which is described by $\hat{h} = \tanh(\omega/2T)\,\hat{\tau}_0$, where $T$ is the temperature and $\hat{\tau}_0$ is an identity matrix.
Thus, we find that
\begin{equation}
    \hat{G}_0^K = (\hat{G}^R_0 - \hat{G}_0^A) \tanh(\omega/2T).
\end{equation}
Another useful symmetry which follows from the definitions of the Green functions~\cite{morten2003a}, is that $\hat{G}^A = (\hat{\tau}_3 \hat{G}^{R} \hat{\tau}_3)^\dagger$, where $\hat{\tau}_3 = \mathrm{diag}(+1, +1, -1, -1)$ is a Pauli matrix in Nambu space.
This lets us further simplify the relation above to
\begin{equation}
    \hat{G}_0^K = (\hat{G}^R_0 - \hat{\tau}_3 \hat{G}_0^{R\dagger} \hat{\tau}_3) \tanh(\omega/2T).
\end{equation}
Using the short-hand notation $\hat{G}^R_{i} = \hat{G}^R_0(\bm{p}_i, \omega)$ for brevity, the above lets us rewrite our equation for the RKKY interaction as
\begin{equation}
    \begin{aligned}
        E_{\text{RKKY}}^{12} = &                                      \\
        \frac{{\pi}\mathcal{J}^2}{{16}} \, \mathrm{Im}
                               & \int d\omega\, \tanh(\omega/2T)      \\
        \times\,               & \int d\bm{p}_{1} \int d\bm{p}_{2} \,
        e^{{i}(\bm{p}_{2} - \bm{p}_{1})\cdot(\bm{R}_2 - \bm{R}_1)}     \\
        \times\,               & \mathrm{Tr}\,
        \Big\{\,
        (\bm{S}_2\cdot\hat{\bm{\sigma}})
        \, \hat{G}_1^R
        \, (\bm{S}_1 \cdot \hat{\bm{\sigma}})
        \, \hat{G}_2^R
        \\ & \phantom{\mathrm{Tr}}\!
        -
        (\bm{S}_2\cdot\hat{\bm{\sigma}})
        \, \hat{G}_1^R
        \, (\bm{S}_1 \cdot \hat{\bm{\sigma}})
        \, \hat{\tau}_3 \hat{G}_2^{R\dagger} \hat{\tau}_3
        \\ & \phantom{\mathrm{Tr}}\!
        +
        (\bm{S}_2\cdot\hat{\bm{\sigma}})
        \, \hat{G}_1^{R}
        \, (\bm{S}_1 \cdot \hat{\bm{\sigma}})
        \,  \hat{\tau}_3 \hat{G}_2^{R\dagger} \hat{\tau}_3
        \\ & \phantom{\mathrm{Tr}}
        -
        (\bm{S}_2\cdot\hat{\bm{\sigma}})
        \, \hat{\tau}_3 \hat{G}_1^{R\dagger} \hat{\tau}_3
        \, (\bm{S}_1 \cdot \hat{\bm{\sigma}})
        \, \hat{\tau}_3 \hat{G}_2^{R\dagger} \hat{\tau}_3
        \,\Big\}.
    \end{aligned}
\end{equation}
Clearly, the second and third terms in the trace cancel.
The remaining terms can be simplified using the fact that ${[\hat{\tau}_3, \hat{\bm{\sigma}}] = 0}$ together with the cyclic trace rule:
\begin{equation}
    \label{eq:almost-done}
    \begin{aligned}
        E_{\text{RKKY}}^{12} = &                                      \\
        \frac{\pi \mathcal{J}^2}{16} \, \mathrm{Im}
                               & \int d\omega\, \tanh(\omega/2T)      \\
        \times\,               & \int d\bm{p}_{1} \int d\bm{p}_{2} \,
        e^{{i}(\bm{p}_{2} - \bm{p}_{1})\cdot(\bm{R}_2 - \bm{R}_1)}     \\
        \times\,               & \mathrm{Tr}\,
        \Big\{\,
        (\bm{S}_2\cdot\hat{\bm{\sigma}})
        \, \hat{G}_1^R
        \, (\bm{S}_1 \cdot \hat{\bm{\sigma}})
        \, \hat{G}_2^R
        \\ & \phantom{\mathrm{Tr}}\!
        -
        (\bm{S}_2\cdot\hat{\bm{\sigma}})
        \, \hat{G}_1^{R\dagger}\!
        \, (\bm{S}_1 \cdot \hat{\bm{\sigma}})
        \, \hat{G}_2^{R\dagger}
        \,\Big\}.
    \end{aligned}
\end{equation}
To further simplify this result, we observe that:
\begin{align*}
     & \left\{\, \int d\bm{p}_1 \int d\bm{p}_2
    e^{{i}(\bm{p}_2 - \bm{p}_1)\cdot(\bm{R}_2-\bm{R}_1)}
    \text{Tr}\left[
        (\bm{S}_2\cdot\hat{\bm{\sigma}})
        \hat{G}_1^R
        (\bm{S}_1 \cdot \hat{\bm{\sigma}})
        \hat{G}_2^R
        \right]
    \,\right\}^*                               \\
     & = \int d\bm{p}_1 \int d\bm{p}_2\,
    e^{{-i}(\bm{p}_2 - \bm{p}_1)\cdot(\bm{R}_2-\bm{R}_1)}\,
    \text{Tr}\left[
        (\bm{S}_2\cdot\hat{\bm{\sigma}})
        \hat{G}_1^R
        (\bm{S}_1 \cdot \hat{\bm{\sigma}})
        \hat{G}_2^R
        \right]^\dagger
    \\
     & = \int d\bm{p}_1 \int d\bm{p}_2\,
    e^{{i}(\bm{p}_1 - \bm{p}_2)\cdot(\bm{R}_2-\bm{R}_1)}\,
    \text{Tr}\left[
    \hat{G}_2^{R\dagger}
    (\bm{S}_1 \cdot \hat{\bm{\sigma}})
    \, \hat{G}_1^{R\dagger}
    (\bm{S}_2\cdot\hat{\bm{\sigma}})
    \right]
    \\
     & = \int d\bm{p}_1 \int d\bm{p}_2\,
    e^{{i}(\bm{p}_1 - \bm{p}_2)\cdot(\bm{R}_2-\bm{R}_1)}\,
    \text{Tr}\left[
    (\bm{S}_2\cdot\hat{\bm{\sigma}})
    \hat{G}_2^{R\dagger}
    (\bm{S}_1 \cdot \hat{\bm{\sigma}})
    \, \hat{G}_1^{R\dagger}
    \right]
    \\
     & = \int d\bm{p}_2 \int d\bm{p}_1\,
    e^{{i}(\bm{p}_2 - \bm{p}_1)\cdot(\bm{R}_2-\bm{R}_1)}\,
    \text{Tr}\left[
    (\bm{S}_2\cdot\hat{\bm{\sigma}})
    \hat{G}_1^{R\dagger}
    (\bm{S}_1 \cdot \hat{\bm{\sigma}})
    \, \hat{G}_2^{R\dagger}
    \right]
\end{align*}
where in the last step we relabeled the momentum variables $\bm{p}_1 \leftrightarrow \bm{p}_2$ (which by definition also implies $\hat{G}^R_1 \leftrightarrow \hat{G}^R_2$).
We now substitute this result into \cref{eq:almost-done}, use the general identity $z - z^* = 2i\,\mathrm{Im}\,z$, and reinstate the definitions of $\hat{G}_i^R$:
\begin{equation}
    \begin{aligned}
        E_{\text{RKKY}}^{12} = &                                      \\
        \frac{{\pi}\mathcal{J}^2}{{8}} \, \mathrm{Im}
                               & \int d\omega\, \tanh(\omega/2T)      \\
        \times\,               & \int d\bm{p}_{1} \int d\bm{p}_{2} \,
        e^{{i}(\bm{p}_{2} - \bm{p}_{1})\cdot(\bm{R}_2 - \bm{R}_1)}     \\
        \times\,               & \mathrm{Tr}\,
        \Big\{\,
        (\bm{S}_2\cdot\hat{\bm{\sigma}})
        \, \hat{G}_0^R(\bm{p}_1, \omega)
        \, (\bm{S}_1 \cdot \hat{\bm{\sigma}})
        \, \hat{G}_0^R(\bm{p}_2, \omega)
        \,\Big\},
    \end{aligned}
\end{equation}

The other contribution $E^{21}_{\text{RKKY}}$ can be obtained by letting $\bm{S}_1 \leftrightarrow \bm{S}_2$ and $\bm{R}_1 \leftrightarrow \bm{R}_2$.
Using the cyclic trace rule, and again relabeling momentum variables $\bm{p}_1 \leftrightarrow \bm{p}_2$, one can then show that $E_{\text{RKKY}}^{12} = E_{\text{RKKY}}^{21}$.
Thus, we can write the final equation for the interaction energy $E_{\text{RKKY}} = E^{12}_{\text{RKKY}} + E^{21}_{\text{RKKY}}$ as
\begin{equation}
    \label{eq:rkky-final}
    \begin{aligned}
        E_{\text{RKKY}} = &                                      \\
        \frac{{\pi}\mathcal{J}^2}{{4}} \, \mathrm{Im}
                          & \int d\omega\, \tanh(\omega/2T)      \\
        \times\,          & \int d\bm{p}_{1} \int d\bm{p}_{2} \,
        e^{{-i}(\bm{p}_{2} - \bm{p}_{1})\cdot(\bm{R}_2 - \bm{R}_1)} \\
        \times\,          & \mathrm{Tr}\,
        \Big\{\,
        (\bm{S}_1 \cdot \hat{\bm{\sigma}})
        \, \hat{G}_0^R(\bm{p}_1, \omega)
        \,(\bm{S}_2\cdot\hat{\bm{\sigma}})
        \, \hat{G}_0^R(\bm{p}_2, \omega)
        \,\Big\}.
    \end{aligned}
\end{equation}
This {result} is valid for general translation-invariant {superconductors} in equilibrium, and can be evaluated {as long as its}  unperturbed Green function $\hat{G}_0^R(\bm{p}, \omega)$ is known.

\section{RKKY interactions in superconductors}\label{sec:specific}
In \cref{sec:general}, we derived \cref{eq:rkky-final} for the RKKY interaction energy $E_{\text{RKKY}}$ in a translation-invariant {superconductor} described by a Green function $\hat{G}_0^R(\bm{p}, \omega)$.
We now specialize to the case where $\hat{G}_0^R$ describes a $p$-wave triplet superconductor, and explore how $E_{\text{RKKY}}$ depends on the symmetries of its order parameter.
First, let us rewrite \cref{eq:rkky-final} as
\begin{equation}
    \label{eq:rkky-split}
    \begin{aligned}
        E_{\text{RKKY}} =
        {\frac{\pi}{4}} \mathcal{J}^2 \, \mathrm{Im}
               & \int d\bm{p}_{1} \int d\bm{p}_{2} \,
        e^{{-i}(\bm{p}_{2} - \bm{p}_{1})\cdot(\bm{R}_2 - \bm{R}_1)}                                                 \\
        \times & \int d\omega\, \tanh(\omega/2T) \, \mathcal{T}(\bm{S}_1, \bm{S}_2, \bm{p}_1, \bm{p}_2, \omega),
    \end{aligned}
\end{equation}
where the function $\mathcal{T}$ at the end refers to the trace
\begin{equation}
    \label{eq:trace}
    \mathcal{T} =
    \mathrm{Tr}\,
    \big[
        (\bm{S}_1 \cdot \hat{\bm{\sigma}})
        \, \hat{G}_1
        \,(\bm{S}_2\cdot\hat{\bm{\sigma}})
        \, \hat{G}_2
        \big],
\end{equation}
and we for brevity use the short-hand notation $\hat{G}_i \equiv \hat{G}^R_0(\bm{p}_i, \omega)$ for the relevant Green function from here on.
The structure of the matrices $\hat{G}_i$ and $\hat{\bm{\sigma}}_i$ in Nambu space is
\begin{align}
    \hat{G}_i                         & =
    \begin{pmatrix}
        G_i         & F_i         \\[1ex]
        \tilde{F}_i & \tilde{G}_i
    \end{pmatrix}, &
    \hat{\bm{\sigma}}                 & =
    \begin{pmatrix}
        \bm{\sigma} &               \\
                    & \bm{\sigma}^*
    \end{pmatrix},
\end{align}
where $\tilde{X}(\bm{p}, \omega) \equiv X^*(-\bm{p}, -\omega)$.
The $2\times2$ spin matrices $G_i$ and $F_i$ are referred to as the normal and anomalous Green functions, respectively.
The normal Green function describes quasi\-particles (electrons and holes), whereas the anomalous Green function describes superconducting correlations (Cooper pairs).
If we substitute these matrices into the bracketed expression in \cref{eq:trace}, and perform an explicit matrix multiplication, we see that the diagonal entries which contribute to the trace are
\begin{widetext}
    \begin{equation}
        (\bm{S}_1\cdot\hat{\bm\sigma}) \hat{G}_1 (\bm{S}_2\cdot\hat{\bm\sigma}) \hat{G}_2 = \\
        \begin{pmatrix}
            (\bm{S}_1\cdot\bm{\sigma}) G_1 (\bm{S}_2\cdot\bm{\sigma}) G_2 +
            (\bm{S}_1\cdot\bm{\sigma}) F_1 (\bm{S}_2\cdot\bm{\sigma}^*) \tilde F_2 &
            \cdots                                                                   \\
            \cdots                                                                 &
            (\bm{S}_1\cdot\bm{\sigma}^*) \tilde G_1 (\bm{S}_2\cdot\bm{\sigma}^*) \tilde G_2 +
            (\bm{S}_1\cdot\bm{\sigma}^*) \tilde F_1 (\bm{S}_2\cdot\bm{\sigma}) F_2 &
        \end{pmatrix}.
    \end{equation}
\end{widetext}
Clearly, the bottom-right block is just the ``tilde conjugate'' of the top-left block.
Thus, taking the trace of this result yields
\begin{align}
    \mathcal{T} & = \mathcal{G} + \tilde{\mathcal{G}} + \mathcal{F} + \tilde{\mathcal{F}},                     \\
    \label{eq:rkky-G}
    \mathcal{G} & \equiv \mathrm{Tr}[(\bm{S}_1\cdot\bm{\sigma}) G_1 (\bm{S}_2\cdot\bm{\sigma}) G_2],           \\
    \label{eq:rkky-F}
    \mathcal{F} & \equiv \mathrm{Tr}[(\bm{S}_1\cdot\bm{\sigma}) F_1 (\bm{S}_2\cdot\bm{\sigma}^*) \tilde{F}_2].
\end{align}
Here, the RKKY interactions mediated by quasiparticles and superconductivity, respectively, are contained in~$\mathcal{G}$ and~$\mathcal{F}$.

Next, let us rewrite the contributions from $\tilde{\mathcal{G}}$ and $\tilde{\mathcal{F}}$ in terms of $\mathcal{G}$~and~$\mathcal{F}$.
Let us first combine \cref{eq:rkky-split} with the definition of tilde conjugation, and then redefine the integration variables $\{\bm{p}_1, \bm{p}_2, \omega\} \rightarrow \{-\bm{p}_1, -\bm{p}_2, -\omega\}$.
This shows us that:
\begin{align*}
             & \int d\bm{p}_{1} \, d\bm{p}_{2} \, d\omega \,
    e^{{-i}(\bm{p}_{2} - \bm{p}_{1})\cdot \delta\bm{R}}
    \tanh(\omega/2T) \,
    \tilde{\mathcal{X}}(\bm{p}_1, \bm{p}_2, \omega)                   \\
    \equiv\, & \int d\bm{p}_{1} \, d\bm{p}_{2} \, d\omega \,
    e^{{-i}(\bm{p}_{2} - \bm{p}_{1})\cdot \delta\bm{R}}
    \tanh(\omega/2T) \,
    \mathcal{X}^*(-\bm{p}_1, -\bm{p}_2, -\omega)                      \\
    =\,      & \int d\bm{p}_{1} \, d\bm{p}_{2} \, d\omega \,
    e^{{+i}(\bm{p}_{2} - \bm{p}_{1})\cdot \delta\bm{R}}
    \tanh(-\omega/2T) \,
    \mathcal{X}^*(\bm{p}_1, \bm{p}_2, \omega)                         \\
    =\,      & -\left\{ \int d\bm{p}_{1} \, d\bm{p}_{2} \, d\omega \,
    e^{{-i}(\bm{p}_{2} - \bm{p}_{1})\cdot \delta\bm{R}}
    \tanh(\omega/2T)  \,
    \mathcal{X}(\bm{p}_1, \bm{p}_2, \omega) \right\}^*,
\end{align*}
where the bracketed expression is simply the corresponding contribution from $\mathcal{X}$.
Since $z - z^* = 2i\,\mathrm{Im}\,z$, we conclude that
\begin{equation}
    \label{eq:rkky-nice}
    \begin{aligned}
        E_{\text{RKKY}} =
        {\frac{1}{2} \pi} \mathcal{J}^2 \, \mathrm{Im}
               & \int d\bm{p}_{1} \int d\bm{p}_{2} \,
        e^{{-i}(\bm{p}_{2} - \bm{p}_{1})\cdot(\bm{R}_2 - \bm{R}_1)} \\
        \times & \int d\omega\, \tanh(\omega/2T) \,
        (\mathcal{G} + \mathcal{F}),
    \end{aligned}
\end{equation}
which shows that we do not need to explicitly calculate the tilde-conjugated contributions $\tilde{\mathcal{G}}$ and $\tilde{\mathcal{F}}$ to evaluate $E_{\text{RKKY}}$.

In the literature, the integrand of the above expression usually contains a Fermi-Dirac distribution function $n_F(\omega)$ rather than a $\tanh$-function, so let us briefly show that the two formulations are equivalent. To do so, we rewrite Eq. (\ref{eq:rkky-nice}) by using that $\tanh(\omega/2T) = 1-2n_F(\omega)$ where $n_F$ is the Fermi-Dirac distribution:
\begin{align}\label{eq:rkkytextbook}
            E_{\text{RKKY}} &=
        -{ \pi} \mathcal{J}^2 \, \mathrm{Im}
                \int d\bm{p}_{1} \int d\bm{p}_{2} \,
        e^{{-i}(\bm{p}_{2} - \bm{p}_{1})\cdot(\bm{R}_2 - \bm{R}_1)} \notag\\
        &\qquad\qquad\times  \int d\omega\, n_F(\omega) \,
        (\mathcal{G} + \mathcal{F}) \notag\\
        &\,+ \frac{1}{2}{ \pi} \mathcal{J}^2 \, \mathrm{Im}
                \int d\bm{p}_{1} \int d\bm{p}_{2} \,
        e^{{-i}(\bm{p}_{2} - \bm{p}_{1})\cdot(\bm{R}_2 - \bm{R}_1)} \notag\\
        &\qquad\qquad\times  \int d\omega\,  \,
        (\mathcal{G} + \mathcal{F}),
\end{align}
The first term is the textbook term. Consider now the second term. The key observation is that all the poles of the retarded Green functions $G(\vecp, \omega)$, $F(\vecp, \omega),$and $ \tilde{F}(\vecp, \omega)$ that enter $\mathcal{G}$ and $\mathcal{F}$ lie in the lower complex energy half-plane. In the absence of a self-energy, this follows since energy enters in the combination $E +i\delta$ with $\delta=0^+$. In the presence of a retarded self-energy $\Sigma$ that enters the expressions for the retarded Green functions, one must have $-\text{Im} \, \Sigma > 0$ since this quantity takes the role of a finite quasiparticle lifetime. From a more precise mathematical viewpoint, $-\text{Im} \, \Sigma > 0$ ensures that the retarded Green function remains analytic in the upper complex energy half-plane, which in turn ensures causality for the solution of field equations with a source term when using the retarded Green function. With this property of the poles of the Green functions, it follows that the last integral in Eq. (\ref{eq:rkkytextbook}) equals zero. This can be seen by closing the contour in the upper halfplane, which contains no poles, using the residue theorem, and noting that the integrand approaches zero sufficiently fast $(1/\omega^2)$ to ensure that the arc contribution to the contour integral vanishes. 

    To make further analytical progress, we need to calculate the contributions $\mathcal G$ and $\mathcal F$ for a $p$-wave triplet superconductor.
    This, in turn, requires that we know the mathematical structures of $G_i$ and~$F_i$.
    In \cref{sec:green-explicit}, we calculate the exact Green function matrix for a general $p$-wave superconductor using block-matrix inversion.
    The results can be written in the form
    \begin{align}
        G_i
            &= g_s(\bm p_i, \omega) + \bm{g}_{p}(\bm p_i, \omega) \cdot \bm\sigma \\
            &\equiv g_{si} + \bm{g}_{pi}, \\[1ex]
        F_i
            &\equiv [\bm{f}_{p}(\bm p_i, \omega) \cdot \bm\sigma] i\sigma_2 \label{eq:anomalous-explicit} \\
            &\equiv ({\bm{f}}_{pi} \cdot \bm\sigma) i\sigma_2.
    \end{align}
    Here, $\bm{f}_p(\bm p, \omega)$ is proportional to the $d$-vector $\bm{d}(\bm p)$, whereas $\bm{g}_p(\bm p, \omega)$ is proportional to the condensate spin expectation value $\bm\mu(\bm p) \equiv i[\bm{d}(\bm p) \times \bm{d}^*(\bm p)]$.
    The exact values of $\{ g_{si}, \bm{g}_{pi}, \bm{f}_{pi} \}$ can be extracted from \cref{eq:green-g-exact,eq:green-f-exact}.

    Let us first consider the quasiparticle contribution~$\mathcal{G}$.
    Substituting the parametrization above into \cref{eq:rkky-G}, we obtain:
    \begin{equation}
        \mathcal{G} = \mathrm{Tr}[(\bm{S}_1 \cdot \bm\sigma) (g_{s1} + \bm{g}_{p1} \cdot \bm\sigma) (\bm{S}_2 \cdot \bm\sigma) (g_{s2} + \bm{g}_{p2} \cdot \bm\sigma)].
    \end{equation}
    Expanding the parentheses above, and using the cyclic trace rule $\mathrm{Tr}[ABC] = \mathrm{Tr}[CAB]$ to reorder some terms, we obtain
    \begin{equation}
    \begin{aligned}
        \mathcal{G}
        &= g_{s1} g_{s2} \mathrm{Tr}[(\bm{S}_1 \cdot \bm\sigma) (\bm{S}_2 \cdot \bm\sigma)] \\
        &\,+ g_{s1} \mathrm{Tr}[(\bm{S}_1 \cdot \bm\sigma) (\bm{S}_2 \cdot \bm\sigma) (\bm{g}_{p2} \cdot \bm\sigma)] \\
        &\,+ g_{s2} \mathrm{Tr}[(\bm{S}_2 \cdot \bm\sigma) (\bm{S}_1 \cdot \bm\sigma) (\bm{g}_{p1} \cdot \bm\sigma)] \\
        &\,+ \mathrm{Tr}[(\bm{S}_1 \cdot \bm\sigma) (\bm{g}_{p1} \cdot \bm\sigma) (\bm{S}_2 \cdot \bm\sigma) (\bm{g}_{p2} \cdot \bm\sigma)].
    \end{aligned}
    \end{equation}
    We now use the following identity for products of Pauli vectors:
    \begin{equation}
        (\bm a \cdot \bm\sigma)(\bm b \cdot \bm\sigma) = (\bm a \cdot \bm b) + i(\bm a \times \bm b)\cdot \bm\sigma.
        \label{eq:pauli-vector-ident}
    \end{equation}
    Repeatedly applying this identity to the equation above, and then evaluating the resulting traces of $2\times2$ matrices, we get
    \begin{equation}
    \begin{aligned}
        \mathcal{G}
        &= 2 g_{s1} g_{s2} (\bm{S}_1 \cdot \bm{S}_2) \\
        &\,+ 2i g_{s1} \bm{g}_{p2} \cdot (\bm{S}_1 \times \bm{S}_2) + 2i g_{s2} \bm{g}_{p1} \cdot (\bm{S}_2 \times \bm{S}_1) \\
        &\,+ 2(\bm{S}_1 \cdot \bm{g}_{p1}) (\bm{S}_2 \cdot \bm{g}_{p2}) - 2(\bm{S}_1 \times \bm{g}_{p1}) \cdot (\bm {S}_2 \times \bm{g}_{p2}).
    \end{aligned}
    \end{equation}
    This result can be simplified as follows.
    For the second line of the right-hand side, we use the antisymmetric property $\bm{S}_1 \times \bm{S}_2 = -\bm{S}_2 \times \bm{S}_1$ to rewrite the term as
    \begin{equation}
    \begin{aligned}
     &2i g_{s1} \bm{g}_{p2} \cdot (\bm{S}_1 \times \bm{S}_2) + 2i g_{s2} \bm{g}_{p1} \cdot (\bm{S}_2 \times \bm{S}_1) \\ = \; & 2i (g_{s1} \bm{g}_{p2} - g_{s2} \bm{g}_{p1}) \cdot (\bm{S}_1 \times \bm{S}_2).
    \end{aligned}
    \end{equation}
    For the last term in $\mathcal{G}$, we can use the quadruple product identity $(\bm a \times \bm b) \cdot (\bm c \times \bm d) = (\bm a \cdot \bm c)(\bm b \cdot \bm d) - (\bm a \cdot \bm d)(\bm b \cdot \bm c)$ to show that
    \begin{equation}
    \begin{aligned}
        & 2(\bm{S}_1 \times \bm{g}_{p1}) \cdot (\bm {S}_2 \times \bm{g}_{p2}) \\
        =\; & 2(\bm{S}_1 \cdot \bm{S}_2) (\bm{g}_{p1} \cdot \bm{g}_{p2}) - 2(\bm{S}_1 \cdot \bm{g}_{p2})(\bm{S}_2 \cdot \bm{g}_{p1}).
    \end{aligned}
    \end{equation}
    Putting together these pieces, we now obtain the result
    \begin{equation}
    \begin{aligned}
        \mathcal{G}
        &= 2 (g_{s1} g_{s2} - \bm{g}_{p1} \cdot \bm{g}_{p2}) (\bm{S}_1 \cdot \bm{S}_2) \\
        &\,+ 2i (g_{s1} \bm{g}_{p2} - g_{s2} \bm{g}_{p1}) \cdot (\bm{S}_1 \times \bm{S}_2) \\
        &\,+ 2(\bm{S}_1 \cdot \bm{g}_{p1}) (\bm{S}_2 \cdot \bm{g}_{p2}) + 2(\bm{S}_1 \cdot \bm{g}_{p2}) (\bm{S}_2 \cdot \bm{g}_{p1}).
    \end{aligned}
    \end{equation}
    If we again use the quadruple product identity, we can see that the expression above can be further simplified via the identity
    \begin{equation}
    \begin{aligned}
        & (\bm{g}_{p1} \times \bm{g}_{p2}) \cdot (\bm S_1 \times \bm S_2) \\
        = \; & (\bm{g}_{p1} \cdot \bm{S}_1) (\bm{g}_{p2} \cdot \bm{S}_2) - (\bm{g}_{p1} \cdot \bm{S}_2) (\bm{g}_{p2} \cdot \bm{S}_1).
    \end{aligned}
    \end{equation}
    After some reordering, our final result for $\mathcal{G}$ can be written
    \begin{equation}
    \begin{aligned}
        \mathcal{G}
        &= 2 (g_{s1} g_{s2} - \bm{g}_{p1} \cdot \bm{g}_{p2}) (\bm{S}_1 \cdot \bm{S}_2) + 4(\bm{g}_{p1} \cdot \bm{S}_1) (\bm{g}_{p2} \cdot \bm{S}_2)\\
        &\,+ 2(ig_{s1} \bm{g}_{p2} - ig_{s2} \bm{g}_{p1} - \bm{g}_{p1}\times\bm{g}_{p2}) \cdot (\bm{S}_1 \times \bm{S}_2).
    \end{aligned}
    \label{eq:mathcal-G-final}
    \end{equation}

    Next, we turn our attention to the condensate contribution~$\mathcal{F}$ to the RKKY interaction in the $p$-wave superconductor.
    Substituting \cref{eq:anomalous-explicit} into \cref{eq:rkky-F}, we obtain
    \begin{equation}
        \mathcal{F} = \mathrm{Tr}[(\bm{S}_1\cdot\bm{\sigma}) (\bm{f}_{p1}\cdot\bm{\sigma}) i\sigma_2 (\bm{S}_2\cdot\bm{\sigma}^*) (\tilde{\bm{f}}_{p2}\cdot\bm{\sigma}^*) i\sigma_2].
    \end{equation}
    We can now use $(\sigma_2)^2 = 1$ to insert an extra pair of $\sigma_2$ matrices between the $(\bm{S}_2 \cdot \bm{\sigma}^*)$ and $(\tilde{\bm{f}}_{p2}\cdot\bm{\sigma}^*)$, and subsequently use $\sigma_2 \bm\sigma \sigma_2 = -\bm{\sigma}^*$ to get rid of all the $\sigma_2$ factors. This yields
    \begin{equation}
        \mathcal{F} = -\mathrm{Tr}[(\bm{S}_1\cdot\bm{\sigma}) (\bm{f}_{p1}\cdot\bm{\sigma}) (\bm{S}_2\cdot\bm{\sigma}) (\tilde{\bm{f}}_{p2}\cdot\bm{\sigma})].
    \end{equation}
    Next, we repeatedly invoke \cref{eq:pauli-vector-ident}, and trace over all the resulting contributions in the same way as for $\mathcal{G}$.
    This yields:
    \begin{equation}
        \mathcal{F} = - 2(\bm{S}_1 \cdot \bm{f}_{p1})(\bm{S}_2 \cdot \tilde{\bm{f}}_{p2}) + 2(\bm{S}_1 \times \bm{f}_{p1}) \cdot (\bm{S}_2 \times \tilde{\bm{f}}_{p2}).
    \end{equation}
    In the last term, we can expand the quadruple product
    \begin{equation}
        (\bm{S}_1 \times \bm{f}_{p1}) \cdot (\bm{S}_2 \times \tilde{\bm{f}}_{p2}) = (\bm{S}_1 \cdot \bm{S}_2)(\bm{f}_{p1} \cdot \tilde{\bm{f}}_{p2}) - (\bm{S}_1 \cdot \tilde{\bm{f}}_{p2})(\bm{S}_2 \cdot \bm{f}_{p1}),
    \end{equation}
    which yields the revised expression
    \begin{equation}
      \begin{aligned}
        \mathcal{F} =\, & -2 (\bm{S}_1 \cdot \bm{f}_{p1})(\bm{S}_2 \cdot \tilde{\bm{f}}_{p2}) \\
              & + 2(\bm{S}_1 \cdot \bm{S}_2)(\bm{f}_{p1} \cdot \tilde{\bm{f}}_{p2}) \\
              &- 2 (\bm{S}_1 \cdot \tilde{\bm{f}}_{p2})(\bm{S}_2 \cdot \bm{f}_{p1}).
      \end{aligned}
    \end{equation}
    Next, we note that another quadruple product could be expanded
    \begin{equation}
        (\bm{f}_{p1} \times \tilde{\bm{f}}_{p2}) \cdot (\bm{S}_1 \times \bm{S}_2) = (\bm{f}_{p1} \cdot \bm{S}_1) (\tilde{\bm{f}}_{p2} \cdot \bm{S}_2) - (\bm{f}_{p1} \cdot \bm{S}_2) (\tilde{\bm{f}}_{p2} \cdot \bm{S}_1).
    \end{equation}
    Substituting this into the above result for $\mathcal{F}$, we finally obtain
    \begin{equation}
    \begin{aligned}
        \mathcal{F}
            &= 2(\bm{f}_{p1} \cdot \tilde{\bm{f}}_{p2}) (\bm{S}_1 \cdot \bm{S}_2) \\
            &-4(\bm{S}_1 \cdot \bm{f}_{p1}) (\bm{S}_2 \cdot \tilde{\bm{f}}_{p2}) \\
            & +2(\bm{f}_{p1} \times \tilde{\bm{f}}_{p2}) \cdot (\bm{S}_1 \times \bm{S}_2).
    \end{aligned}
    \label{eq:mathcal-F-final}
    \end{equation}

    We have now calculated both contributions $\mathcal{G}$ and $\mathcal{F}$ to $E_{\text{RKKY}}$ in a system with $p$-wave triplet superconductivity.
    We can now substitute the results in \cref{eq:mathcal-G-final,eq:mathcal-F-final} into \cref{eq:rkky-nice} to obtain the following equation for the RKKY interaction energy $E_{\text{RKKY}}$ in a general $p$-wave superconductor:
    \begin{equation}
        \begin{aligned}
            E&_{\text{RKKY}} = \\
            & \pi \mathcal{J}^2 \, \mathrm{Im} \int d\bm{p}_{1} \int d\bm{p}_{2} \, \int d\omega\, \tanh(\omega/2T)
            e^{{-i}(\bm{p}_{2} - \bm{p}_{1})\cdot(\bm{R}_2 - \bm{R}_1)} \\
            & \times \Big\{\,
                 (g_{s1} g_{s2} - \bm{g}_{p1} \cdot \bm{g}_{p2} + \bm{f}_{p1} \cdot \tilde{\bm{f}}_{p2}) (\bm{S}_1 \cdot \bm{S}_2) \\
                 &\quad + 2(\bm{g}_{p1} \cdot \bm{S}_1) (\bm{g}_{p2} \cdot \bm{S}_2) - 2(\bm{S}_1 \cdot \bm{f}_{p1}) (\bm{S}_2 \cdot \tilde{\bm{f}}_{p2}) \\
                &\quad + (ig_{s1} \bm{g}_{p2} - ig_{s2} \bm{g}_{p1} - \bm{g}_{p1}\times\bm{g}_{p2} + \bm{f}_{p1} \times \tilde{\bm{f}}_{p2}) \cdot (\bm{S}_1 \times \bm{S}_2)
            \Big\}
        \end{aligned}
    \end{equation}
    Note that except for the complex exponential prefactor, each term in the integrand above contains only one factor
    \begin{equation}
        \alpha(\bm{p}_1, \omega) \in \{ g_{s1}, \bm{g}_{p1}, \bm{f}_{p1} \}
    \end{equation}
    that depends on the first momentum variable $\bm{p}_1$.
    We see that the integral over $\bm{p}_1$ is simply a Fourier transform of this factor:
    \begin{equation}
        \frac{1}{(2\pi)^3} \int d\bm{p}_1 \, e^{i\bm{p}_1 \cdot \bm{\delta}} \alpha(\bm{p}_1, \omega) = \alpha(\bm{\delta}, \omega),
    \end{equation}
    where we for brevity defined the variable $\bm{\delta} \equiv \bm{R}_2 - \bm{R}_1$.
    Each term also contains only one factor $\beta(\bm{p}_2, \omega)$ depending on $\bm{p}_2$, and this part also takes the form of a Fourier transform:
    \begin{equation}
            \frac{1}{(2\pi)^3} \int d\bm{p}_2 \, e^{i\bm{p}_2 \cdot (-\bm{\delta})} \beta(\bm{p}_2, \omega) = \beta(-\bm{\delta}, \omega).
            \label{eq:fourier-beta-def}
    \end{equation}
    Here, $\beta(\bm{\delta}, \omega)$ is then the Fourier transform of either $g_{s}(\bm{p},\omega)$, $\bm{g}_{p}(\bm{p}, \omega)$, or $\tilde{\bm{f}}_p(\bm{p}, \omega)$.
    But the derivations in \cref{sec:green-explicit} indicate that $g_s$ and $\bm{g}_p$ must be even functions of~$\bm{p}$, which in turn makes their Fourier transforms even functions of~$\bm{\delta}$.
    This follows directly from \cref{eq:green-g-exact}: If we note that the $d$-vector is by definition an odd function of~$\bm{p}$, and each term that arises in both the numerator and denominator contains only even powers of this $d$-vector, then the overall term must necessarily be an even function of~$\bm{p}$.
    Based on this information, we can judge that $g_s(-\bm{\delta}, \omega) = g_s(\bm{\delta}, \omega)$ and $\bm{g}_p(-\bm{\delta}, \omega) = \bm{g}_p(\bm{\delta}, \omega)$.
    On the other hand, $\bm{f}_p(\bm{p}, \omega)$---and thus $\tilde{\bm{f}}_p(\bm{p}, \omega)$---must be an odd function of momentum.
    This can also be seen explicitly in \cref{eq:green-f-exact}: Each term contains an odd number of $\bm{d}(\bm{p})$ factors, where the $d$-vector itself is again an odd function of momentum.
    The Fourier transform of an odd function is always an odd function, which implies that $\bm{f}_p(-\bm{\delta}, \omega) = -\bm{f}_p(\bm{\delta}, \omega)$.
    Based on the discussion above, we conclude that $E_{\text{RKKY}}$ is then:
    \begin{equation}
        \begin{aligned}
            E&_{\text{RKKY}} = \\
            & 64\pi^7 \mathcal{J}^2 \, \mathrm{Im} \int d\omega\, \tanh(\omega/2T) \\
            & \times \Big\{\,
                 (g_{s}^2 - \bm{g}_{p}^2 - \bm{f}_{p} \cdot \tilde{\bm{f}}_{p}) (\bm{S}_1 \cdot \bm{S}_2) \\
                 &\quad + 2(\bm{g}_{p} \cdot \bm{S}_1) (\bm{g}_{p} \cdot \bm{S}_2) + 2(\bm{f}_{p} \cdot \bm{S}_1) (\tilde{\bm{f}}_{p} \cdot \bm{S}_2) \\
                &\quad + (ig_{s} \bm{g}_{p} - ig_{s} \bm{g}_{p} - \bm{g}_{p}\times\bm{g}_{p} + \bm{f}_{p} \times \tilde{\bm{f}}_{p}) \cdot (\bm{S}_1 \times \bm{S}_2)
            \Big\}
        \end{aligned}
    \end{equation}
    Here, we write $g_s \equiv g_s(\bm{\delta}, \omega), \bm{g}_p \equiv \bm{g}_p(\bm{\delta}, \omega), \bm{f}_p \equiv \bm{f}_p(\bm{\delta}, \omega)$ for the real-space components of the Green function.
 
    Surprisingly, we can show that \emph{all} the DMI contributions cancel at this point.
    Clearly, $ig_s \bm{g}_p - ig_s \bm{g}_p = 0$ and $\bm{g}_p \times \bm{g}_p = 0$ vanish for trivial reasons.
    The remaining candidate $\bm{f}_p \times \tilde{\bm{f}}_p$ requires a bit more explanation.
    If we explictly write out this DMI contribution to the RKKY interaction, it becomes
    \begin{equation}
    \begin{aligned}
       E_{\text{DMI}} &= \bm{D} \cdot (\bm{S}_1 \times \bm{S}_2), \\
       \bm{D} &\sim \mathrm{Im} \int\limits_{-\infty}^{+\infty} d\omega\,  \tanh(\omega/2T) \, [\bm{f}_p(\bm{\delta}, \omega) \times \tilde{\bm{f}}_p(\bm{\delta}, \omega)].
    \end{aligned}
    \end{equation}
    Next, we note that $\tilde{\bm{f}}_p(\bm{\delta}, \omega) = \bm{f}_p^*(\bm{\delta}, -\omega)$.
    This can be seen by Fourier transforming the definition $\tilde{\bm{f}}_p(\bm{p}, \omega) = \bm{f}_p^*(-\bm{p}, -\omega)$.
    Since $\tanh(\omega/2T)$ is real-valued, we can rewrite the above as
    \begin{align}
       \bm{D} &\sim \int\limits_{-\infty}^{+\infty} d\omega\, \tanh(\omega/2T) \, \mathrm{Im} [\bm{f}_p(\bm{\delta}, \omega) \times \bm{f}_p^*(\bm{\delta}, -\omega)].
    \end{align}
    Since $\tanh(\omega/2T)$ is an odd function of energy~$\omega$, this integral can only yield a finite contribution to~$\bm{D}$ if $\mathrm{Im}[\cdots]$ also contains an odd-in-$\omega$ contribution.
    But since $\mathrm{Im}[\bm{a} \times \bm{b}] = -\mathrm{Im}[\bm{b} \times \bm{a}] = \mathrm{Im}[\bm{b}^* \times \bm{a}^*]$ for general complex vectors, we conclude that
    \begin{equation}
      \mathrm{Im} [\bm{f}_p(\bm{\delta}, \omega) \times \bm{f}_p^*(\bm{\delta}, -\omega)] =
      \mathrm{Im} [\bm{f}_p(\bm{\delta}, -\omega) \times \bm{f}_p^*(\bm{\delta}, \omega)],
    \end{equation}
    which proves that the $\mathrm{Im}[\cdots]$ expression is in fact an even function of energy~$\omega$.
    This concludes the proof that the remaining DMI contribution is in fact zero.
    Note that the same proof does not hold for e.g. the $\bm{f}_p \cdot \tilde{\bm{f}}_p$ contribution to Heisenberg interaction.
    This is because $\mathrm{Im}[\bm{a} \cdot \bm{b}] = \mathrm{Im}[\bm{b} \cdot \bm{a}] = -\mathrm{Im}[\bm{b}^* \cdot \bm{a}^*]$ yields an expression $\mathrm{Im}[\cdots]$ that is explicitly odd-in-$\omega$, and which thus can produce a finite contribution to the RKKY interaction when inserted into the integral above.
  
    Our final analytical result for the DMI interaction in a translation-invariant $p$-wave superconductor is then
    \begin{equation}
        \begin{aligned}
            E_{\text{RKKY}} &= \\
            64\pi^7 &\mathcal{J}^2 \, \mathrm{Im} \int d\omega\, \tanh(\omega/2T) \\
            \times \Big\{\,& (g_{s}^2 - \bm{g}_{p}^2 - \bm{f}_{p} \cdot \tilde{\bm{f}}_{p}) (\bm{S}_1 \cdot \bm{S}_2) \\
               &+ 2(\bm{g}_{p} \cdot \bm{S}_1) (\bm{g}_{p} \cdot \bm{S}_2) + 2(\bm{f}_{p} \cdot \bm{S}_1) (\tilde{\bm{f}}_{p} \cdot \bm{S}_2)
            \Big\}.
        \end{aligned}
    \end{equation}
    In a non-magnetic normal metal, we only have the $g_s$ contribution, which leads to a pure Heisenberg interaction proportional to $\bm{S}_1 \cdot \bm{S}_2$.
    In a unitary $p$-wave superconductor, we also get finite values for $\bm{f}_p$ and $\tilde{\bm{f}}_p$, which (i)~gives rise to a new Ising interaction and (ii)~modulates the existing Heisenberg interaction.
    Finally, in non-unitary $p$-wave superconductors, there is in addition a finite magnetic term $\bm{g}_p \sim \bm{d} \times \bm{d}^*$, which can influence both the Heisenberg and Ising interactions.
    On the other hand, for infinite and translation-invariant superconductors, we find that the DMI interaction between the spins is in fact zero.
    This is in contrast to the numerical results presented in the main article, which shows that one in a finite sample in fact can obtain a finite DMI term for non-unitary $p$-wave superconductors.
    We refer to \cref{sec:green-explicit} for explicit expressions for the Green function components $\{g_s, \bm{g}_p, \bm{f}_p\}$.
 
\section{Calculation of the Green function}
\label{sec:green-explicit}
    To evaluate the equation for $E_{\text{RKKY}}$ in \cref{sec:specific}, we required an explicit expression for the Green function matrix of a $p$-wave superconductor.
    In this appendix, we derive the required result via explicit matrix inversion.

    Consider a general $p$-wave superconductor.
    In momentum space, such a material is well-described by the $4\times4$ Hamiltonian
    \begin{equation}
        \hat{H}(\bm p) =
        \begin{pmatrix}
            H(+\bm p) & \Delta(+\bm p) \\
            -\Delta^*(-\bm p) & -H^*(-\bm p) \\
        \end{pmatrix},
    \end{equation}
    where $\bm p$ is the momentum degree of freedom.
    For simplicity, we take $H(\bm p)$ to be the Hamiltonian of a normal metal with chemical potential~$\mu$ but no spin-dependent properties,
    \begin{align}
        H(+\bm p) &= H^*(-\bm p) = \xi(\bm p), &
        \xi(\bm p) &\equiv \frac{p^2}{2m} - \mu.
    \end{align}
    For brevity, we do not explicitly write out identity matrices in this derivation, so $\xi(\bm p)$ should e.g. be interpreted as $\xi(\bm p) \sigma_0$ here.
    As in the main article, we use the standard $d$-vector parametrization for the gap matrix of a $p$-wave superconductor,
    \begin{equation}
        \Delta(+\bm p) = -\Delta(-\bm p) = [\bm{d}(\bm p)\cdot\bm\sigma] i\sigma_2.
    \end{equation}
    The symmetries of $H(\bm p)$ and $\Delta(\bm p)$ let us write $\hat H(\bm p)$ as
    \begin{equation}
        \hat{H}(\bm p) =
        \begin{pmatrix}
            \xi(\bm p) & \Delta(\bm p) \\
            \Delta^*(\bm p) & -\xi(\bm p)
        \end{pmatrix}.
    \end{equation}

    We are now interested in calculating the Green function
    \begin{align}
        \hat{G}(\bm p, \omega)
        &\equiv [\omega - \hat{H}]^{-1} \\
        &= \begin{pmatrix}
            \omega - \xi(\bm p) & -\Delta(\bm p) \\
            -\Delta^*(\bm p) & \omega + \xi(\bm p)
        \end{pmatrix}^{-1} \\
        &\equiv \begin{pmatrix}
            G(\bm p, \omega) & F(\bm p, \omega) \\
            \tilde{F}(\bm p, \omega) & \tilde{G}(\bm p, \omega)
        \end{pmatrix}
    \end{align}
    Note that at this point, $\omega$ is kept as a general complex-valued parameter.
    To obtain the retarded Green function $\hat{G}^R(\bm p, \omega)$ considered in the other appendices, one should let $\omega \rightarrow \omega + i0^+$ after the following derivations have been completed.

    The Green function can be found via block-matrix inversion.
    Specifically, if the diagonal blocks and their Schur complements are invertible, then the inverse of a block matrix can be written
    \begin{equation}
        \begin{pmatrix}
            \bm A & \bm B \\
            \bm C & \bm D
        \end{pmatrix}^{-1}
        =
        \begin{pmatrix}
            \bm S^{-1} & -\bm S^{-1} \bm B \bm D^{-1} \\[1ex]
            -\bm D^{-1} \bm C \bm S^{-1} & \phantom{.}\bm D^{-1} \bm C \bm S^{-1} \bm B \bm D^{-1} + \bm D^{-1} \\
        \end{pmatrix},
    \end{equation}
    where $\bm S \equiv \bm A - \bm B {\bm D}^{-1} \bm C$ is the Schur complement of block $\bm D$.
    Applied to the Green function above, we find that the inverse of the normal Green function $G$ is simply given by
    \begin{equation}
        G^{-1}(\bm p, \omega) = \omega - \xi(\bm p) - \frac{\Delta(\bm p) \Delta^*(\bm p)}{\omega + \xi(\bm p)}.
    \end{equation}
    Inserting the $d$-vector parametrization of $\Delta(\bm p)$, and using the Pauli vector identity $(\bm a \cdot \bm \sigma)(\bm b \cdot \bm \sigma) = (\bm a \cdot \bm b) + i(\bm a \times \bm b) \cdot \bm\sigma$, we find that $\Delta \Delta^* = (\bm d \cdot \bm d^*) +i(\bm d \times \bm d^*)\cdot \bm\sigma$.
    Thus, the above result can be written in terms of the $d$-vector as
    \begin{equation}
        G^{-1} = \frac{ \omega^2 - \xi^2 - |\bm d|^2 - i(\bm d \times \bm d^*) \cdot \bm\sigma}{\omega + \xi},
    \end{equation}
    where the dependence on $\bm p$ has been left out for brevity.
    Next, it is straight-forward to verify that a $2\times2$ matrix parametrized via Pauli matrices has a matrix inverse given by
    \begin{equation}
        (u_0 + \bm u \cdot \bm\sigma)^{-1} = \frac{u_0 - \bm u \cdot \bm\sigma}{u_0^2 - \bm u^2}.
    \end{equation}
    This lets us invert $G^{-1}$ to obtain the normal Green function
    \begin{equation}
        G = \frac{(\omega + \xi) [\omega^2 - \xi^2 -|\bm d|^2 +i(\bm d \times \bm d^*)\cdot \bm\sigma]}{(\omega^2 -\xi^2 -|\bm d|^2)^2 + (\bm d\times \bm d^*)^2}.
    \end{equation}
    The denominator can be slightly simplified by noting that $(\bm d \times \bm d^*)^2 = (\bm d \times \bm d^*) (\bm d \times \bm d^*) = -(\bm d \times \bm d^*) (\bm d^* \times \bm d) = - |\bm d\times \bm d^*|^2$, where we used the identity $\bm a \times \bm b = -\bm b \times \bm a$. Thus:
    \begin{equation}
        \label{eq:green-g-exact}
        G = \frac{(\omega + \xi) [\omega^2 - \xi^2 -|\bm d|^2 +i(\bm d \times \bm d^*)\cdot \bm\sigma]}{(\omega^2 -\xi^2 -|\bm d|^2)^2 - |\bm d\times \bm d^*|^2}.
    \end{equation}
    From this exact result for the normal Green function $G(\bm p, \omega)$ of a $p$-wave superconductor, we see that we can in general write
    \begin{equation}
        G(\bm p, \omega) = g_s(\bm p, \omega) + \bm{g}_p(\bm p, \omega) \cdot \bm\sigma,
    \end{equation}
    where the spin-dependent part of the normal Green function $\bm{g}_p \sim i(\bm{d} \times \bm{d}^*) \sim \bm\mu$ is found to be proportional to the spin expectation value $\bm{\mu}$ of the superconducting condensate.
    The results above are also consistent with $G(\bm p, \omega)$ being an even function of momentum~$\bm p$ in a centrosymmetric system.
    This can be seen by using that $\xi(\bm p)$ is even while $\bm d(\bm p)$ is odd in $\bm p$.

    Let us now consider the anomalous Green function $F(\bm p, \omega)$.
    Using the formula for block-matrix inversion, we see that
    \begin{equation}
    \begin{aligned}
        F(\bm p, \omega)
        &= G(\bm p, \omega) \frac{\Delta(\bm p)}{\omega + \xi} \\
        &= \frac{\omega^2 - \xi^2 -|\bm d|^2 +i(\bm d \times \bm d^*)\cdot \bm\sigma}{(\omega^2 -\xi^2 -|\bm d|^2)^2 - |\bm d\times \bm d^*|^2} (\bm d \cdot \bm\sigma)i\sigma_2.
    \end{aligned}
    \end{equation}
    According to the identity $(\bm a \cdot \bm\sigma)(\bm b \cdot \bm\sigma) = (\bm a \cdot \bm b) + i(\bm a \times \bm b)\cdot \bm\sigma$, the product $[(\bm d \times \bm d^*) \cdot \bm \sigma][(\bm d \cdot \bm \sigma)] = (\bm d \times \bm d^*) \cdot \bm d + i[(\bm d \times \bm d^*) \times \bm d] \cdot \bm\sigma$.
    The first of these contributions can be rewritten as $(\bm d \times \bm d) \cdot \bm d^*$ using the cyclic rule for triple products, and this is clearly zero.
    Thus, the anomalous Green function for a $p$-wave superconductor has been shown to be
    \begin{align}
        F(&\bm p, \omega) \nonumber\\
        &= \frac{\omega^2 - \xi^2 - |\bm d|^2}{(\omega^2 -\xi^2 -|\bm d|^2)^2 - |\bm d\times \bm d^*|^2} (\bm d \cdot \bm\sigma) i\sigma_2 \label{eq:green-f-exact} \nonumber \\
          &\phantom{==}- \frac{1}{(\omega^2 -\xi^2 -|\bm d|^2)^2 - |\bm d\times \bm d^*|^2} [((\bm d \times \bm d^*) \times \bm d) \cdot \bm\sigma] i\sigma_2 \\
          &=\, \frac{\omega^2 - \xi^2}{(\omega^2 -\xi^2 -|\bm d|^2)^2 - |\bm d\times \bm d^*|^2} (\bm d \cdot \bm\sigma) i\sigma_2  \nonumber \\
          &\phantom{==}- \frac{1}{(\omega^2 -\xi^2 -|\bm d|^2)^2 - |\bm d\times \bm d^*|^2} [(\bm{d}\cdot\bm{d})(\bm{d}^{*}\cdot\bm{\sigma})i\sigma_2]
    \end{align}
    where the last transition used the vector triple product identity $(\bm{a} \times \bm{b})\times\bm{c} = (\bm{a}\cdot\bm{c}) \bm{b} - (\bm{b}\cdot\bm{c})\bm{a}$ to write the result compactly.

    This result is consistent with $F(\bm p, \omega)$ being an odd function of momentum, as expected for a $p$-wave superconductor.
    From this result, we see that we can write
    \begin{equation}
        F(\bm p, \omega) = [\bm{f}_p(\bm p, \omega) \cdot \bm\sigma] i\sigma_2,
    \end{equation}
    where \cref{eq:green-f-exact} shows that $\bm f_{p}$ has contributions proportional to $\bm d$ and $(\bm d \times \bm d^*)\times \bm d$.
    For all $d$-vectors considered in this article, $(\bm d \times \bm d^*) \times \bm d$ is either zero (unitary superconductors) or proportional to $\bm d$ (the non-unitary state we considered), in which case $\bm{f}_p(\bm p, \omega) \sim \bm d(\bm p)$.
    We will therefore use this form for the derivations in \cref{sec:specific}.

\bibliography{references}
\end{document}